\documentclass[aps,pra,twocolumn,superscriptaddress,notitlepage,nofootinbib,longbibliography]{revtex4-2}
\usepackage{mathrsfs}
\usepackage{epsfig}
\usepackage{graphicx}
\usepackage{amsfonts}
\usepackage[figuresright]{rotating}
\usepackage{amssymb}
\usepackage{amsmath}
\usepackage{dcolumn}
\usepackage{bm}
\usepackage{dutchcal}
\usepackage{xcolor}
\usepackage{color}
\usepackage{braket}
\usepackage{units}
\usepackage{xspace}

\usepackage{bbold}
\definecolor{mypink3}{cmyk}{0, 0.7808, 0.4429, 0.1412}
\definecolor{mypink1}{rgb}{0.858, 0.188, 0.478}
\definecolor{mypink2}{RGB}{219, 48, 122}
\usepackage[colorlinks=true, allcolors=mypink1]{hyperref}
\usepackage[fleqn]{mathtools}

\usepackage[shortlabels]{enumitem}
\newcommand{\ECNU}{Quantum Institute for Light and Atoms, State Key Laboratory of Precision Spectroscopy, Department of Physics, School of Physics and Electronic Science, East China Normal University, Shanghai 200062, China}
\newcommand{\HB}{Shanghai Branch, Hefei National Laboratory, Shanghai 201315, China}
\newcommand{\SPA}{
	School of Physics and Astronomy, and Tsung-Dao Lee Institute, Shanghai Jiao Tong University, Shanghai 200240, China}
\newcommand{\SR}{Shanghai Research Center for Quantum Sciences, Shanghai 201315, China}
\newcommand{\CIC}{Collaborative Innovation Center of Extreme Optics, Shanxi University, Taiyuan, Shanxi 030006, China}

\begin{document}
\title{Dissipative quantum Fisher information for a general Liouvillian parameterized process}

\author{Jia-Xin Peng}
\affiliation{\ECNU}
\affiliation{\HB}
\author{Baiqiang Zhu}
\affiliation{\ECNU}
\affiliation{\HB}
\author{Weiping Zhang}
\affiliation{\HB}
\affiliation{\SPA}
\affiliation{\SR}
\affiliation{\CIC}
\author{Keye Zhang}
\email{kyzhang@phy.ecnu.edu.cn}
\affiliation{\ECNU}
\affiliation{\HB}
\date{\today}

\begin{abstract}
The dissipative quantum Fisher information (DQFI) for a  dynamic map with a general parameter in an open quantum system is investigated, which can be regarded as an analog of the quantum Fisher information (QFI) in the Liouville space. We first derive a general dissipative generator in the Liouville space, and based on its decomposition form, find the DQFI stems from two parts. One is the dependence of eigenvalues of the Liouvillian supermatrix on the estimated parameter, which shows a linear dependence on time. The other is the variation of the eigenvectors with the estimated parameter. The relationship between this part and time presents rich characteristics, including harmonic oscillation, pure exponential gain and attenuation, as well as exponential gain and attenuation of oscillatory type, which depend specifically on the properties of the Liouville spectrum. This is in contrast to that of the conventional generator, where only oscillatory dependencies are seen. Further, we illustrate the theory through a toy model: a two-level system with spin-flip noise. Especially, by using the DQFI, we demonstrated that the exceptional estimation precision cannot be obtained at the Liouvillian exceptional point. 
\end{abstract}
\maketitle
\section{Introduction} \label{I}
Accurate estimation of the values of parameters characterizing an underlying physical setting has important applications in a wide range of scientific fields, which has promoted the rapid development of parameter estimation theory \cite{degen2017quantum,giovannetti2011advances,liu2020quantum}. Quantum Fisher information (QFI), “being at the heart of quantum parameter estimation theory”, is the supremum of the classical Fisher information (CFI) \cite{fisher1925theory} and its inverse is used to quantify the lower bound of mean-square error of unbiased estimator about the unknown parameter \cite{helstrom1969quantum,holevo2011probabilistic,paris2009quantum}. This indicates that the larger QFI of the unknown parameter, the higher the estimation precision that may be achieved. Interestingly enough, QFI plays a key role in other aspects besides characterizing estimation precision, such as measuring the statistical distinguishability between adjacent quantum states \cite{wootters1981statistical,braunstein1994statistical}, the witness of quantum correlations \cite{hyllus2012fisher,toth2012multipartite,hradil2019quantum,banerjee2018connecting} and non-Markovian effects \cite{toth2014quantum,de2017dynamics,lu2010quantum,song2015quantum}, characterizing quantum phase transition \cite{wang2014quantum,marzolino2017fisher,ma2009fisher,sun2010fisher}, bound quantum speed limit \cite{taddei2013quantum,deffner2017quantum,gessner2018statistical,del2013quantum,beau2017nonlinear} and so on.

There is no doubt that the most pivotal task in the research of quantum parameter estimation is to calculate the QFI of unknown parameters. Considering the convenience of the solution, early researchers mainly focused on estimating the overall factor of Hamiltonian, that is, $\widehat{H}_{\theta}=\theta \widehat{H}$ \cite{lang2013optimal,jarzyna2012quantum,bollinger1996optimal,joo2011quantum,joo2012quantum,zhang2013quantum,jing2014quantum}, in which $\theta$ is the parameter to be estimated. For instance, the phase estimation in the interferometer models. However, in many physical models, the estimated parameters do not necessarily appear as the overall factor of Hamiltonian \cite{pang2014quantum,jing2015maximal}. A prominent example is a two-level system in a magnetic field, which is described by the Hamiltonian $\widehat{H}_\theta =B\left[ \cos (\theta)\widehat{\sigma}_{x}+\sin (\theta )\widehat{\sigma}_{z}\right]$ \cite{hou2020minimal,hou2021super}. The direction angle of magnetic field $\theta$, as the parameter to be estimated, is not the overall factor of Hamiltonian in the current case. In 2014, Pang and Brun studied the quantum metrology problem of this kind of general Hamiltonian parameter in detail~\cite{pang2014quantum}. The results show that QFI is proportional to the variance of the conventional generator of local unknown parameter translation. More interestingly, they found that QFI originated from two parts based on the decomposition form of the generator: one is the dependence of the eigenvalues of the Hamiltonian on the parameter to be estimated, which is proportional to time $t$; the other is the dependence of the eigenvectors of the Hamiltonian on the estimated parameter, which oscillates with time $t$. This is a major breakthrough in the study of parameter estimation in closed quantum systems, launching a blast of upsurge in quantum parameter estimation based on the generator method~\cite{jing2015maximal,hou2020minimal,hou2021super,chu2021dynamic,pang2017optimal,hou2021zero,yang2022multiparameter,qvarfort2021optimal,schneiter2020optimal,cheng2022quantum}.

Further progress in parameter estimation theory stemmed from incorporating environment-induced decoherence processes since a real physical system inevitably interacts with the surrounding environment~\cite{demkowicz2015quantum,haase2016precision}. 
Compared with the closed system, the introduction of environmental noise significantly increases the difficulty of analytical solutions, which hinders us from deeply understanding the influence of noise on estimation precision. 
The breakthrough point to solve this difficulty is that Fujiwara~\emph{et al.} proposed for the first time to apply quantum channels technology to the metrological problem~\cite{fujiwara2008fibre}. The resulting general framework quantifying the estimation precision bounds in the decoherence models including the simulation-based methods \cite{matsumoto2010metric,demkowicz2012elusive}, the purification-based \cite{escher2011general} and the channel-extension-based methods \cite{kolodynski2013efficient}. Furthermore, the variational method can also be utilized to calculate the fundamental metrological bounds for some decoherence models \cite{knysh2014true}. In particular, S. Alipour~\emph{et al.} considered a variant of QFI for computational simplicity \cite{alipour2014quantum}, whose essential idea is to transform the estimation problem from Hilbert space to Hilbert-Schmidt space (Liouville space) for processing.  From then on, we named the variant QFI proposed by S. Alipour~\emph{et al.} “dissipative quantum Fisher information" (DQFI). Interestingly, the DQFI is proportional to the covariance between $\vec{L}_{\theta }$ and $\vec{L}_{\theta }^{\dag }$, here $\vec{L}_{\theta }$ refers to the matrix representation of Liouvillian superoperator in the Liouville space. So that the connection between estimation precision and properties of the underlying dynamics of the open system is established, which is difficult to realize under the conventional QFI framework due to the complexity brought by superoperators. It is worth mentioning that, compared to other methods, the DQFI gives the correct scaling of error with unprecedented simplicity for open systems as well as the constant factor for capturing precision scales is more accurate \cite{demkowicz2012elusive,escher2011general,alipour2014quantum}. Subsequently, Benatti \emph{et al.} applied DQFI to the parameter estimation of $N$-particle two-mode bosonic systems with dissipation and enjoyed great success \cite{benatti2014dissipative}. In fact, as early as 2005, the problem of revisting quantum concepts in Liouville space has been presented in Ref.~\cite{sarandy2005adiabatic}, where the authors generalized the concept of adiabaticity to the realm of open systems.

On the other hand, there are several variants of QFI, depending on the differences in definition method, such as QFIs based on (symmetric, right, left) logarithmic derivatives \cite{helstrom1969quantum,liu2020quantum,yuen1973multiple,fujiwara1995quantum}, and the paradigmatic QFI closely related to the skew information \cite{wigner1963information,zhong2013fisher}. Here the first three are mainly applied in the quantum metrology field, while the last one plays an important role in the discrimination of quantum states \cite{zhong2013fisher}.
In recent years, the above variant QFIs have developed rapidly in different application scenarios, but the relevant theories of DQFI have not been further developed since it was proposed. 
We noticed that for the sake of mathematical simplicity, the original theory of DQFI has made an ideal assumption that the parameter to be estimated is an overall factor of Liouvillian supermatrix \cite{alipour2014quantum}, namely $\vec{L}_{\theta}=\theta \vec{L}$ where $\theta$ is the parameter to be estimated, or a little more relaxed, that $[ \vec{L}_{\theta },\partial_{\theta} \vec{L}_{\theta}] = 0$. The former implies that the estimated parameter simply rescales the evolution time of the open quantum system, the physical models studied in Refs. \cite{alipour2014quantum} and \cite{benatti2014dissipative} are limited to this situation. 
By comparison, the latter assumption means that the parameter can simply be an overall factor of the coherent part or the dissipative part of the Liouvillian supermatrix but the two parts of dynamics have to be commutative.

Nevertheless, two common as well as exemplary physical scenarios do not meet the above assumptions: 
(i)~The estimated parameter is not an overall factor of Hamiltonian~\cite{jing2015maximal,hou2020minimal,hou2021super,yang2022multiparameter, qvarfort2021optimal,schneiter2020optimal,cheng2022quantum}. 
For example, as we mentioned before, a two-level system in the magnetic field and affected by noise at the same time, with the magnetic field direction as the estimated parameter. 
(ii)~For open systems where coherent and dissipative dynamics are non-commutative (non-phase-covariant dynamics), which is characteristic of most kinds of noises \cite{wu2020quantum,zhang2021non,sha2022continuous,haase2018fundamental,ferialdi2017momentum,obada2012effects}.
For instance, a photon-loss Jaynes-Cummings model with the Rabi frequency as the parameter to be estimated. 
As a consequence, the assumption that $[\vec{L}_{\theta },\partial_{\theta} \vec{L}_{\theta}] = 0$ may limits the application scenarios and development of DQFI.  
Considering that DQFI is a new avenue and exhibits great power in dealing with quantum parameter estimation problems in open systems. It is thus urgent and significant to extend DQFI theory to general Liouvillian parameterized processes, i.e., $[ \vec{L}_{\theta },\partial_{\theta} \vec{L}_{\theta}] \neq 0$. 
This necessarily provides some theoretical guidance for researchers (to circumvent the involved superoperators) who want to utilize DQFI to deal with parameter estimation problems or study other interesting physical phenomena in general dissipative systems.  

Here, inspired by Pang \emph{et al.}'s method for systems without dissipation \cite{pang2014quantum}, we first derive an effective generator for a dynamic map with a general parameter in the Liouville space for dissipative systems. Further, we express this generator by eigenvalues and eigenvectors of the Liouvillian supermatrix, and find that the DQFI originates from the following two parts: one is the dependence of eigenvalues on parameters to be estimated, which is linear dependence on time $t$; the second is the variation of the eigenvectors with the estimated parameters and the relation between this part and time $t$ is closely related to the gaps of Liouville spectrum.  Remarkably, we find that the non-unitary encoding process induced by the dissipative generator can be transformed into two unitary encoding processes plus the contribution of the commutator between the two. We then apply the theory to a concrete example: a two-level system with spin-flip noise (corresponding to non-phase-covariant dynamics), taking the transition frequency as the estimated parameter. The dynamic evolution of DQFI over time is carefully studied for different decay rates, and the physical mechanism behind it is explained in detail. 

It is worth mentioning that in recent years using Hamiltonian exceptional point (HEP) to improve measurement precision (owing to susceptibility diverges at the exceptional point) has been widely studied in quantum metrology \cite{wiersig2014enhancing,wiersig2016sensors}. 
Nevertheless, the influence of quantum noise on exceptional-point-based sensors has been controversially debated in some theoretical studies \cite{wiersig2020prospects,chen2019sensitivity,langbein2018no,lau2018fundamental}. 
To demonstrate the remarkable efficacy of DQFI in dissipative quantum metrology, we also explored the behavior of QFIs around the Liouvillian exceptional point (LEP). The result clearly shows that the estimation precision near the LEP is not significantly enhanced. 

To facilitate the reading, in Table~\ref{T1} we concisely list some important physical terms as well as corresponding abbreviations and symbols. Note that in order to better distinguish, we add “conventional" in front of the familiar QFI and its related words, and “dissipative" in front of the variant version proposed by S. Alipour \emph{et al.}.
\begin{table}[tbh]
	\label{T1}
	\centering
	\caption{Symbol, abbreviations and corresponding full names used in the paper. }
	\label{1}
	\begin{tabular}{lc}
		\hline
		\hline
		Symbol: Full name   &  Abbreviation  \\ 
		\hline
		\hline
		$F_{c}$: Conventional  classical Fisher information   &  CCFI   \\
		$F_{q}$: Conventional quantum	Fisher information &  CQFI  \\
		Conventional  quantum   Cram\'er-Rao bound &  CQCRB\\
		$\widehat{\mathfrak{M}}$: Conventional  symmetric logarithmic derivative &  CSLD \\
		$\widetilde{F}$: Dissipative  quantum Fisher information  &  DQFI \\
		Dissipative  quantum   Cram\'er-Rao bound & DQCRB \\
    	$\widetilde{\mathfrak{M}}$: Dissipative  symmetric logarithmic derivative &  DSLD \\
		Positive-operator valued measure & POVM\\
		Hamiltonian exceptional point & HEP\\
		Liouvillian exceptional point & LEP\\
		\hline
		\hline
	\end{tabular}
\end{table} 

The paper is organized as follows. In Sec.~\ref{II}, we introduce the related knowledge of parameter estimation and conventional quantum Fisher information  (CQFI). In Sec.~\ref{III}, the dissipative generator for a general dynamical map and corresponding DQFI are derived in the Liouville space, and their related properties are discussed in detail. Sec.~\ref{IV}, we demonstrate the theory by using a specific dissipative two-level system and focused on studying the dynamic evolution of DQFI then compared it with CQFI. The last section concludes this paper. To ensure the integrity of this paper, we present three appendices, in which we review some basic properties of superoperator and vectorization (Appendix \ref{AA}), review the main aspects of non-Hermitian matrix and its adjoint (Appendix \ref{AB}), and give the spectral decomposition formulas for calculating DQFI and dissipative symmetric logarithmic derivative (DSLD) (Appendix \ref{AC}).

\section{A review of basic theories}\label{II}
\begin{figure}[hbt!]
	\centering
	\includegraphics[width=0.99\linewidth]{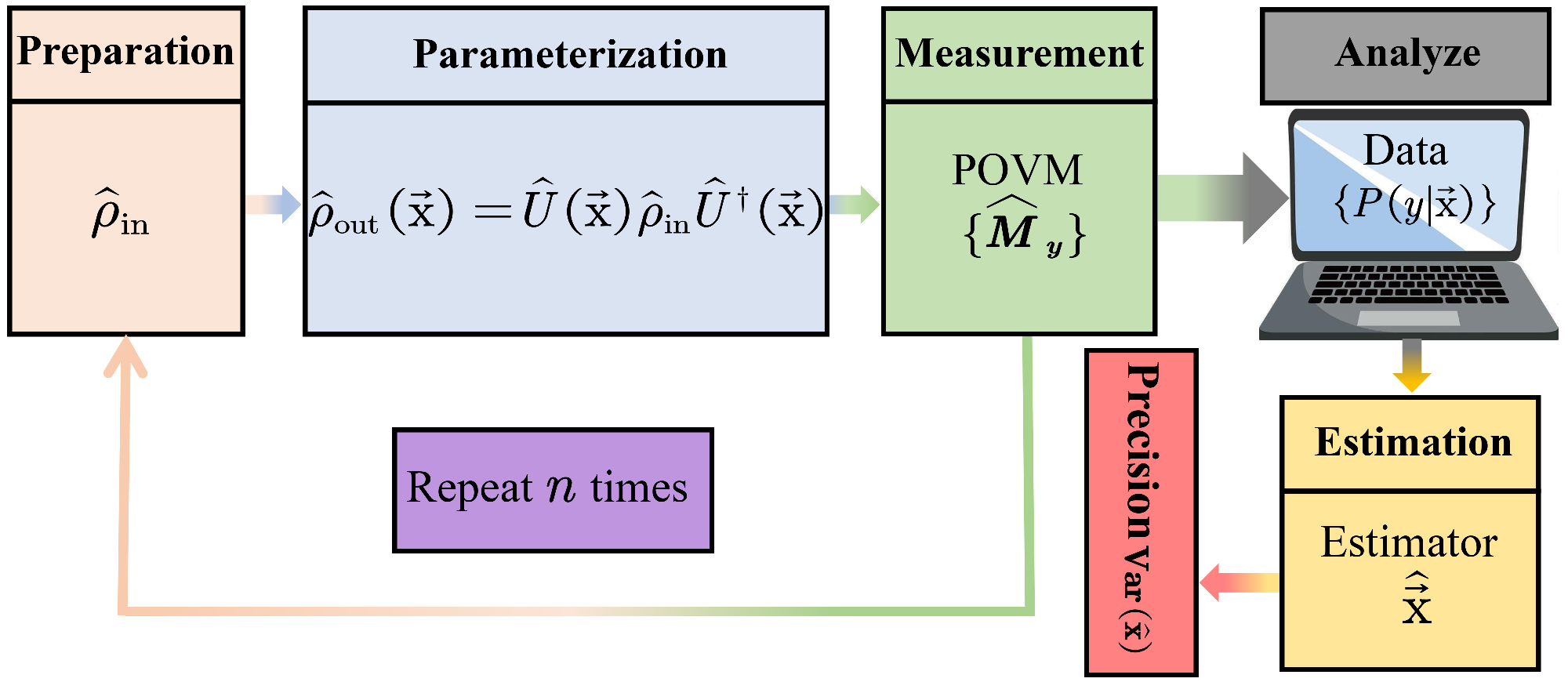}
	\caption{Schematic diagram of complete parameter estimation protocol, which contains four
		steps:~(i)~Preparation of quantum probe state; (ii)~Parameterization processes; (iii)~POVM measurements; (iv)~Estimation.}
	\label{fig0}
\end{figure}

To facilitate the comparison with DQFI, let's start by reviewing the essential features and formulas of CQFI. Suppose $\mathcal{H}$ is a Hilbert space that corresponds to a physical system described by Hamiltonian $\widehat{H}\left( \vec{\mathbf{x}}\right) $ with the
unknown vector of parameters $\vec{\mathbf{x}}:=\left[ \text{x}_{0},\text{x}_{1},...,\text{x}_{n}\right] ^{\text{T}}$. 
In order to infer the unknown parameters information, one needs to employ the means of parameter estimation. 
A typical parameter estimation protocol usually consists of the following four steps, as shown in Fig.~\ref{fig0} \cite{degen2017quantum,giovannetti2011advances,liu2020quantum}: 
(i)~Preparation of quantum probe state $\widehat{\rho} _{\text{in}}$. 
(ii)~Parameterization processes: let the probe state $\widehat{\rho} _{\text{in}}$ evolve under the dynamic mapping $\widehat{U}\left( \vec{\mathbf{x}}\right) =e^{-i\widehat{H}\left( \vec{\mathbf{x}}\right) t}$ where we set $\hbar=1$ in this paper, so that encoding information about the unknown parameter to the output state, that is $\widehat{\rho} _{\text{out}}\left( \vec{\mathbf{x}}\right) =\widehat{U}\left( \vec{\mathbf{x}}\right)
\widehat{\rho} _{\text{in}}\widehat{U}^{\dag }\left( \vec{\mathbf{x}}\right) $. 
(iii)~Construct a set
of positive-operator valued measure (POVM) $\{ \bm{\widehat{M}_{y}}\} $ that satisfy positive semidefinite as well as completeness,
and then measuring the output state to obtain the conditional probability distribution of
the reading results $\{P(y|\vec{\mathbf{x}})=\text{Tr}[\widehat{\rho}_{\text{out}}(\vec{\mathbf{x}}) \bm{\widehat{M}_{y}}]\} $ (given by the Born rule), here $y$ denotes the measurement outcome. Let us emphasize that in one time complete estimation protocol, in order to obtain the conditional probability distribution, one needs to perform $\nu $ times measurements on the identical parameter-dependent quantum state. 
Remember not to confuse $\nu $ with the number of protocol repetitions $n$ mentioned later.
(iv)~The unbiased estimator $\widehat{\vec{\mathbf{x}}}:=\left[ \widehat{\text{x}}_{0},\widehat{\text{x}}_{1},...,\text{x}_{n} %
\right] ^{\text{T}}$ is constructed, which is a statistic that satisfies $\vec{\mathbf{x}}=\langle \widehat{\vec{\mathbf{x}}}\rangle$ \cite{helstrom1969quantum,holevo2011probabilistic,paris2009quantum}. 
Further, the obtained data are processed and analyzed by statistical techniques to infer indirectly the information on parameters of interest. 
The above steps qualitatively describe the main process of parameter estimation. 
In order to quantitatively describe the estimation precision of the unknown parameters, one needs to resort to the conventional quantum Cram\'er-Rao bound (CQCRB), which is the most renowned metrological tool in quantum estimation theory.  
In the single-parameter estimation scenario, the CQCRB inequality satisfies the following relationship~\cite{degen2017quantum,giovannetti2011advances,liu2020quantum,fisher1925theory,helstrom1969quantum,holevo2011probabilistic,paris2009quantum}
\begin{equation}
\label{e1}
\text{Var}\left( \widehat{\text{x}}\right) \geq \frac{1}{nF_{c}(\{\bm{\widehat{M}_{y}}%
	\})}\geq \frac{1}{nF_{q}}, 
\end{equation}
with
\begin{subequations}
	\begin{eqnarray}
	\text{Var}\left(\widehat{\text{x}}\right)  &=&\sum\limits_{y}P(y|\text{x})\left( 
	\widehat{\text{x}}-\text{x}\right) ^{2}, \\
	F_{c}(\{\bm{\widehat{M}_{y}}\}) &=&\sum\limits_{y}\frac{1}{P(y|\text{x})}\left[
	\partial _{\text{x}}P(y|\text{x})\right] ^{2}, \\
    F_{q} &=&\text{Tr}[\widehat{\rho}_{\text{out}}(\text{x}) \widehat{\mathfrak{M}}_{%
		\text{x}}^{2}],  \label{e2c}
	\end{eqnarray}%
\end{subequations}
where $\text{Var}\left( \text{\^{x}}\right)$ is the mean-square error for unbiased estimator $\widehat{\text{x}}$, characterizing the average deviation of the estimated value from the true; $F_{c}(\{\bm{\widehat{M}_{y}}\}\}$ stands for conventional  classical Fisher information (CCFI), which depends on the specific measurement strategies; 
$F_{q}$ represents the CQFI, which relies only on the Hamiltonian and initial state of the system as well as already optimized over all theoretically admissible estimators and measurement schemes; 
$n$ denotes  the repetition of the estimation protocol or equivalently, the number of independent probes; 
$\widehat{\mathfrak{M}}_{\text{x}}$ being the conventional  symmetric logarithmic derivative (CSLD) operator for unknown parameter \text{x}, which is a Hermitian operator satisfying $2\partial _{\text{x}}\widehat{\rho}_{\text{out}}\left( \text{x}\right) =%
\widehat{\rho}_{\text{out}}\left( \text{x}\right) \widehat{\mathfrak{M}}_{\text{x}}+\widehat{\mathfrak{M}}_{\text{x}}\widehat{\rho}_{\text{out}}\left( \text{x}\right)$. According to Eq.~(\ref{e1}), in order to obtain higher estimation precision, the Fisher information or $n$ should be as big as possible. Crucially, $F_{c}=F_{q}$ for single parameter estimation scenario when $\{ \bm{\widehat{M}_{y}}\}$ is constructed from the eigenbasis of the CSLD~\cite{liu2020quantum,kolodynski2013efficient}, namely optimal measurement strategy.

At present, various formulas for calculating CQFI have been derived based on the definition. Here, we
mainly introduce the method of calculating CQFI based on
conventional generator, which is closely related to our current work. Let $\widehat{H}\left( \theta
\right) $ be the Hamiltonian of the system we study with the initial state $%
\widehat{\rho}_{\text{in}}$ (without parameters to be estimated), and $\theta $
is the estimated parameter. For a closed system, the evolved state under the action of time-independent Hamiltonian is $\widehat{\rho}_{%
	\text{out}}\left( \theta,t \right) =\widehat{U}\left( \theta  \right) \widehat{\rho}_{%
	\text{in}}\widehat{U}^{\dag }\left( \theta  \right) $, where $\widehat{U}\left( \theta 
\right) =e^{-i\widehat{H}\left( \theta \right) t}$ is parameter-dependent time evolution operator. One can define a Hermitian operator representing the local conventional generator of parameter
translation with respect to $\theta$~\cite{pang2014quantum,giovannetti2006quantum,brody2013information},
\begin{equation}
\widehat{h}_{\theta }=i[\partial _{\theta }\widehat{U}\left( \theta \right)
] \widehat{U}^{\dag }\left( \theta \right). \label{e3}
\end{equation}
Based on the definitions of
CQFI [see Eq.~(\ref{e2c})] and $\widehat{h}_{\theta }$, the CQFI of parameter $\theta $ for the closed system reads~\cite{liu2020quantum},
\begin{eqnarray}
F_{q}\left( \theta ,t\right)  &=&\sum\limits_{i=1}^{S}\frac{\left[ \partial
	_{\theta }p_{i}\left( \theta \right) \right] ^{2}}{p_{i}\left( \theta
	\right) }+\sum\limits_{i=1}^{S}4p_{i}\left( \theta \right) \left\langle \psi
_{i}\left( \theta \right) \right\vert \widehat{h}_{\theta }^{2}|\psi _{i}\left(
\theta \right) \rangle   \notag \\
&&-\sum\limits_{i,j=1}^{S}\frac{8p_{i}p_{j}}{p_{i}+p_{j}}|\left\langle \psi
_{i}\left( \theta \right) \right\vert \widehat{h}_{\theta }|\psi _{j}\left(
\theta \right) \rangle |^{2}. \label{e4}
\end{eqnarray}%
Here $|\psi _{i}\left( \theta \right) \rangle $ and $p_{i}\left( \theta
\right) $ are the $i$th eigenstate and corresponding eigenvalue of $\widehat{\rho%
}_{\text{out}}\left( \theta ,t\right) $, respectively. $S$ is the support
dimension of $\widehat{\rho}_{\text{out}}\left( \theta ,t\right) $. Suppose that
the initial state of system is a pure state, Eq.~(\ref{e4}) can be reduced
to 
\begin{eqnarray}
F_{q}\left( \theta ,t\right)  &=&4\text{Cov}_{|\psi \left( \theta \right)
	\rangle }(\widehat{h}_{\theta },\widehat{h}_{\theta })  \notag \\
&=&4\left\langle \psi \left( \theta \right) \right\vert \Delta \widehat{h}%
_{\theta }^{2}|\psi \left( \theta \right) \rangle =4\langle \Delta \widehat{h}%
_{\theta }^{2}\rangle .  \label{e5}
\end{eqnarray}%
This shows that, for the case of pure state the CQFI is proportional to the variance of local conventional
generator $\widehat{h}_{\theta }$. 
Further, the uncertainty relation 
\begin{equation}
\label{e6aa}
\langle \Delta \widehat{\theta}^{2}\rangle \langle \Delta \widehat{h}_{\theta
}^{2}\rangle \geq 1/4
\end{equation}
can be obtained for a single metrology protocol by combining Eqs.~(\ref{e1})
and~(\ref{e5}). Obviously, in order to obtain the ultimate precision limit, $%
\langle \Delta \widehat{h}_{\theta }^{2}\rangle $ needs to be maximized. One can
find this is achieved under a special supposition state 
\begin{equation}
\left\vert \Psi \left( \theta \right) \right\rangle =\dfrac{1}{\sqrt{2}}%
\left[ |\eta _{\text{max}}(\widehat{h}_{\theta })\rangle +e^{i\phi }|\eta _{%
	\text{min}}(\widehat{h}_{\theta })\rangle \right] ,  \label{e7aa}
\end{equation}%
where $|\eta _{\text{max}}(\widehat{h}_{\theta })\rangle $ [$|\eta _{\text{min}}(\widehat{h}_{\theta })\rangle$] is the eigenstate corresponding to the maximum
(minimum) eigenvalue $\eta _{\text{max}}(\widehat{h}_{\theta })$ [$\eta _{\text{min}}(\widehat{h}_{\theta })$] of $\widehat{h}_{\theta }$ \cite{braunstein1994statistical,giovannetti2006quantum,braunstein1996generalized} and $\phi$ is an arbitrary phase. Substituting $\left\vert \Psi \left( \theta \right) \right\rangle $ into
Eq.~(\ref{e5}), the maximal CQFI reads 
\begin{equation}
F_{q}^{\text{max}}\left( \theta ,t\right) =\left[ \eta _{\text{max}}(\widehat{h}%
_{\theta })-\eta _{\text{min}}(\widehat{h}_{\theta })\right] ^{2}.  \label{e6}
\end{equation}%
Note that this maximum value is given under the situation that the optimal
initial state of the system is without unknown parameters. The optimal
initial state can be deduced by a time reversal, e.g., $\left\vert \Psi _{%
	\text{in}}\right\rangle _{\text{opt}}=\widehat{U}^{\dagger }(\theta )\left\vert
\Psi \left( \theta \right) \right\rangle $. If the optimal initial state
contains parameters to be estimated, it needs to be prepared adaptively
based on accumulated data \cite{chen2020fluctuation}.

Particularly, when the estimated parameter is an overall factor of
Hamiltonian, e.g., $\widehat{H}\left( \theta \right) =\theta \widehat{H}$, the
generator $\widehat{h}_{\theta }=\widehat{H}t$ and Eq.~(\ref{e5}) can be further
reduced to 
\begin{equation}
F_{q}\left( \theta ,t\right) =4t^{2}\left\langle \psi _{\text{in}%
}\right\vert \Delta \widehat{H}^{2}|\psi _{\text{in}}\rangle ,  \label{e7}
\end{equation}%
where $\left\vert \psi _{\text{in}}\right\rangle $ denotes the initial state
of the system. From the above formula, one can predict that when the
Hamiltonian is time independent, the CQFI is proportional to $t^{2}$ for the
closed system, namely achieving the Heisenberg precision \cite%
{pang2014quantum,hou2021super}. 
Nevertheless, when $\widehat{H}(\theta )\neq \theta \widehat{H}$, such as the example mentioned in the introduction earlier $\widehat{H}_\theta =B\left[ \cos (\theta)\widehat{\sigma}_{x}+\sin (\theta )\widehat{\sigma}_{z}\right]$, the maximal CQFI $F_{q}^{\text{max}}\left( \theta, t\right)=4$sin$^{2}(Bt)$ \cite{pang2014quantum,yuan2015optimal}, indicating that more time may even lead to worse estimation precision. This exhibits a significant difference between the parameter to be estimated as an overall factor and not as an overall factor. This is one of the motivations for the research in this paper.

\section{Dissipative generator for a general dynamical map and corresponding DQFI in the Liouville space}\label{III}     
In this section, we will present the derivation process of a dissipative generator for a dynamic map with a general parameter and the corresponding DQFI in the Liouville space. 
Firstly, one needs to prepare the mathematical framework required to derive the dissipative generator for open quantum systems. Let the time evolution of an open quantum system described by $M$-dimensional Hilbert space obeys the convolutionless master equation \cite{breuer2002theory,rivas2012open}  
\begin{equation}
\label{e8}
\partial _{t}\widehat{\rho}_{s}\left( \theta,t\right)  =\widehat{\mathcal{L}}(\theta,t)\left[
\widehat{\rho}_{s}\left( \theta,t\right) \right], \\ 
\end{equation} 
with
\begin{align}
\widehat{\mathcal{L}}(\theta ,t)\left[ \bullet \right] &: =-i[\widehat{H}\left(
\theta ,t\right) ,\bullet ]  \notag \\
& +\frac{1}{2}\sum_{k}\gamma _{k}\left( \theta ,t\right) \left( [\widehat{\Gamma}%
_{k},\bullet \widehat{\Gamma}_{k}^{\dag }]+[\widehat{\Gamma}_{k}\bullet ,\widehat{\Gamma}%
_{k}^{\dag }]\right) .
\end{align}
Here $\widehat{\mathcal{L}}(\theta,t)$ is the Liouvillian superoperator, in which $\widehat{H}\left( \theta,t\right) $ is time-dependent internal Hamiltonian, the substitution symbol ``$\bullet$'' is transformed operator (e.g., density matrices), and quantum jump operator $\widehat{\Gamma}_{k}$ is associated
with a dissipation quantum channel occurring at the rate $\gamma
_{k}\left(\theta, t\right) $. Note that the estimated parameter $\theta$ can appear in the Hamiltonian $\widehat{H}\left(
\theta,t\right) $, or in $\gamma_{k}\left(\theta, t\right) $, or even in both. It's easy to get 
$\widehat{\rho}_{s}\left( \theta,t\right) =\bm{\widehat{\mathcal{T}}} e^{\int\nolimits_{0}^{t}\widehat{\mathcal{L}}(\theta,\tau)d\tau} \widehat{\rho}_{s}\left( 0\right)$
by performing formal integration on Eq.~(\ref{e8}), where $\bm{\widehat{\mathcal{T}}}$
is a chronological time-ordering operator, but the disadvantage of this density matrix-formed integral is that the physical interpretation is
extremely difficult since it contains a superoperator. 

In light of this, one can vectorize the master equation (\ref{e8}) using the Hilbert-Schmidt expression of the density matrix \cite{casagrande2021analysis,minganti2019quantum,caves1999quantum}, namely rewrite the $M\times M$ density matrix $\widehat{%
	\rho}_{s}\left( \theta,t\right) $ into a $M^{2}\times 1$ column vector, i.e., $\left\vert \widehat{\rho}_{s}\left( \theta,t\right) \rangle \right\rangle =[ 
\begin{array}{cccc}
\widehat{\rho}_{s}\left( 1,:\right)  & \widehat{\rho}_{s}\left( 2,:\right)  & \cdots 
& \widehat{\rho}_{s}\left( M,:\right) 
\end{array}%
] ^{\text{T}}$, where $\widehat{\rho}_{s}\left( i,:\right) $ represents the $i$th row of the density matrix, and the double bracket notation indicates that we are working in the Hilbert-Schmidt (Liouville) space of state vectors. 
The evolution equation that $\left\vert \widehat{\rho}%
_{s}\left( \theta,t\right) \rangle \right\rangle $ satisfies can be obtained
after straightforward algebra
\begin{equation}
\label{e9}
\partial _{t}\left\vert \widehat{\rho}_{s}\left( \theta,t\right) \rangle
\right\rangle =\vec{L}(\theta,t)\left\vert \widehat{\rho}_{s}\left( \theta,t\right)
\rangle \right\rangle ,
\end{equation}%
with
\begin{eqnarray}
\label{e9c}
&&\vec{L}(\theta ,t)=-i\left[ \widehat{H}\left( \theta ,t\right) \otimes \mathbb{%
	1}_{M}-\mathbb{1}_{M}\otimes \widehat{H}\left( \theta ,t\right) ^{^{\text{T}}}%
\right]    \\
&&+\sum_{k}\gamma _{k}\left( \theta ,t\right) \left[ \widehat{\Gamma}_{k}\otimes 
\widehat{\Gamma}_{k}^{\ast }-\dfrac{1}{2}(\widehat{\Gamma}_{k}^{\dag }\widehat{\Gamma}%
_{k}\otimes \mathbb{1}_{M}+\mathbb{1}_{M}\otimes \widehat{\Gamma}_{k}^{\text{T}}%
\widehat{\Gamma}_{k}^{\ast })\right],\notag
\end{eqnarray}
which is a Sch\"{o}dinger-liked equation without superoperator. 
Here operator $\vec{L}(\theta,t)$ denotes $M^{2}\times M^{2}$-dimensional non-Hermitian matrix formed by vectorization of Liouvillian superoperator $\widehat{\mathcal{L}}(\theta,t)$, and $\mathbb{1}_{M}$ represents $M\times M$-dimensional identity matrix. Then the quantum state of the system can be obtained by the integral $\left\vert \widehat{\rho}_{s}\left( \theta,t\right) \rangle \right\rangle =%
\bm{\widehat{\mathcal{T}}} e^{^{\int\nolimits_{0}^{t}\vec{L}%
		(\theta,\tau)d\tau}}\left\vert \widehat{\rho}_{s}\left( 0\right) \rangle \right\rangle $,
where $\left\vert \widehat{\rho}_{s}\left( 0\right) \rangle \right\rangle $ is the initial state of the system in the Liouville space.  
This is the advantage of dealing with open systems through vectorization technology, which circumvents the complexity caused by the abstract superoperator.   
Notice that  $%
\left\vert \widehat{\rho}_{s}\left( \theta,t\right) \rangle \right\rangle $ is not
necessarily normalized owing to  non-unitary dynamics, and its corresponding normalized pure state can be
defined as  \cite{alipour2014quantum}
\begin{equation}
\left\vert \widehat{\rho}_{s}\left( \theta,t\right) \rangle \right\rangle _{N}
=\left\vert \widehat{\rho}_{s}\left( \theta,t\right) \rangle \right\rangle /\sqrt{%
	 \langle \left\langle \widehat{\rho}_{s}\left( \theta,t\right) |\widehat{%
		\rho}_{s}\left( \theta,t\right) \right\rangle \rangle  }.
\end{equation}
Here and hereinafter, the right subscript “$N$" of the state refers to the normalized mark.

We are now ready to derive the dissipative generator for general Liouvillian parameterized processes in the Liouville space. 
According to Eq.~(\ref{e9}), one obtain $%
\widetilde{\rho}_{s}\left( \theta ,t\right) =\widetilde{U}_{\theta }\widetilde{\rho}%
_{s}\left( 0\right) \widetilde{U}_{\theta }^{\dag }$, where $\widetilde{\rho}%
_{s}\left( \theta ,t\right) =\left\vert \widehat{\rho}_{s}\left( \theta
,t\right) \rangle \right\rangle \langle \left\langle \widehat{\rho}_{s}\left(
\theta ,t\right) \right\vert $ and $\widetilde{U}_{\theta }=\bm{\widehat{\mathcal{T}}} e^{^{\int\nolimits_{0}^{t}\vec{L}(\theta ,\tau)d\tau}}$ denotes the
non-Hermitian evolution operator in the Liouville space. $\widetilde{\rho}_{s}\left( \theta
,t\right) $ is not normalized, which can be processed in the latter calculation of DQFI.
Hence the response of $\widetilde{\rho}_{s}\left( \theta ,t\right) $ to small deviations 
around a given the estimated parameter  $\theta $ can be represented by the dissipative generator of the
local parameter translation from $\widetilde{\rho}_{s}\left( \theta ,t\right) $
to $\widetilde{\rho}_{s}\left( \theta +d_{\theta} ,t\right) $, where $d_{\theta} $ denotes
an infinitesimal quantity $\left( d_{\theta} /\theta \rightarrow 0\right) $. 
On the other hand, 
\begin{eqnarray}
\label{e11} 
\widetilde{\rho}_{s}\left( \theta +d_{\theta} ,t\right)  &=&\widetilde{U}_{\theta
	+d_{\theta} }\widetilde{\rho}_{s}\left( 0\right) \widetilde{U}_{\theta +d_{\theta} }^{\dag
}  \notag \\
&\approx &\left[ \widetilde{U}_{\theta }+\left( \partial _{\theta }\widetilde{U}%
_{\theta }\right) d_{\theta} \right] \widetilde{\rho}_{s}\left( 0\right) \left[ 
\widetilde{U}_{\theta }^{\dag }+\left( \partial _{\theta }\widetilde{U}_{\theta
}^{\dag }\right) d_{\theta} \right]   \notag \\
&=&\left[ \mathbb{1}_{M^2}+\left( \partial _{\theta }\widetilde{U}_{\theta }\right) \widetilde{U}%
_{\theta }^{-1}d_{\theta} \right] \widetilde{U}_{\theta }\widetilde{\rho}_{s}\left(
0\right) \widetilde{U}_{\theta }^{\dag }  \notag \\
&&\left[\mathbb{1}_{M^2}+\left( \widetilde{U}_{\theta }^{\dag }\right) ^{-1}\left( \partial
_{\theta }\widetilde{U}_{\theta }^{\dag }\right) d_{\theta} \right]  \notag \\
&\approx &e^{-i \widetilde{\Xi} _{\theta }d_{\theta }}\widetilde{\rho}_{s}\left( \theta
,t\right) e^{i \widetilde{\Xi} _{\theta }^{\dag }d_{\theta }},
\end{eqnarray}%
where 
\begin{equation} 
\label{e12} 
\widetilde{\Xi} _{\theta }=i \left( \partial _{\theta }\widetilde{U}_{\theta }\right) \widetilde{U}%
_{\theta }^{-1},\,\,\widetilde{\Xi} _{\theta }^{\dag }=-i \left( \widetilde{U}_{\theta }^{\dag
}\right) ^{-1}\left( \partial _{\theta }\widetilde{U}_{\theta }^{\dag }\right).   
\end{equation}%
In Eq.~(\ref{e11}), we assumed that $\widetilde{U}_{\theta +d_{\theta} }\approx \widetilde{U}%
_{\theta }+\left( \partial _{\theta }\widetilde{U}_{\theta }\right) d_{\theta} $ by
Taylor expansion under the condition $d_{\theta} /\theta \rightarrow
0$. $\widetilde{\Xi} _{\theta }$ denotes the dissipative generator of the dynamical
map $\widetilde{U}_{\theta }$ along parameter $\theta$ in the Liouville space, which is generally
non-Hermitian owing to the existence of dissipation, unlike the conventional generator $\widehat{h}_{\theta }$ in the closed system. This is the first core formula of our current work. 
When the noise does not exist, $\widetilde{\Xi}_{\theta }$ returns to a Hermitian operator. This is because $\vec{L}^{\dag }(\theta ,t)=-\vec{L}%
(\theta ,t)$ holds in the absence of noise, resulting in  $\widetilde{U}_{\theta
}\widetilde{U}_{\theta }^{\dag }=\mathbb{1}_{M^2}$. Then combined with $\partial _{\theta
}( \widetilde{U}_{\theta }\widetilde{U}_{\theta }^{\dag }) =0$, it can be
proven that  $\widetilde{\Xi}_{\theta }=\widetilde{\Xi}_{\theta }^{\dag }$ by Eq.~(\ref{e12}). In particular, $\widetilde{U}_{\theta }$ may not be invertible in some special cases, such as when some eigenvalues of $\vec{L}(\theta,t)$ go to minus infinity. 
In these cases, one can introduce the Moore-Penrose
pseudoinverse of $\widetilde{U}_{\theta }$ \cite{penrose1955generalized}, i.e., 
\begin{equation}
\widetilde{U}_{\theta
}^{-1}=\lim\limits_{\delta \rightarrow 0}\left[ \widetilde{U}_{\theta }^{\dag
}\left( \widetilde{U}_{\theta }\widetilde{U}_{\theta }^{\dag }+\delta \mathbb{1}_{M^2}\right) ^{-1}%
\right],
\end{equation}
which always exists.

Therefore, with similar mathematical techniques, the definition of CQFI and CSLD
in the closed systems can also be directly extended to the Liouville space, that
is \cite{alipour2014quantum}
\begin{equation}
\label{e14}
\widetilde{F}\left( \theta ,t\right)=\text{Tr}\left[ \widetilde{\rho}_{s}\left( \theta ,t\right) _{N}\widetilde{\mathfrak{M}}_{\theta }^{2}\right] ,
\end{equation}%
with 
\begin{equation}
\widetilde{\rho}_{s}\left( \theta ,t\right) _{N} =\left\vert \widehat{\rho}%
_{s}\left(\theta,t\right) \rangle \right\rangle \left\langle \langle \widehat{\rho}%
_{s}\left( \theta,t\right) \right\vert /\text{Tr}[ \widehat{\rho}_{s}\left(
\theta,t\right) ^{2}] ,
\end{equation}
and
\begin{equation}
\partial _{\theta }\widetilde{\rho}_{s}\left( \theta ,t\right) _{N} =\dfrac{1}{%
	2}\left[ \widetilde{\rho}_{s}\left( \theta ,t\right) _{N}\widetilde{\mathfrak{M}}_{\theta }+%
\widetilde{\mathfrak{M}}_{\theta }\widetilde{\rho}_{s}\left( \theta ,t\right) _{N}\right] .
\end{equation}
Here $\widetilde{F}\left( \theta ,t\right)$ and $\widetilde{\mathfrak{M}}_{\theta }$ are respectively  DQFI and
DSLD under the extended description. In particular, the density matrix of the system can always be
regarded as a pure state in the Liouville space, hence 
\begin{eqnarray}
\label{e15}
\widetilde{\mathfrak{M}}_{\theta } &\equiv &2\partial _{\theta }\widetilde{\rho}_{s}\left(
\theta ,t\right) _{N}  \notag \\
&=&2[\partial _{\theta }\left\vert \widehat{\rho}_{s}\left( \theta ,t\right)
\right\rangle \rangle _{N}\left\langle \langle \widehat{\rho}_{s}\left( \theta
,t\right) \right\vert   \notag \\
&&+\left\vert \widehat{\rho}_{s}\left( \theta ,t\right) \right\rangle \rangle
_{N}\partial _{\theta N}\langle \left\langle \widehat{\rho}_{s}\left( \theta
,t\right) \right\vert ].
\end{eqnarray}%
In order to calculate
DQFI, $\partial _{\theta }\widetilde{\rho}_{s}\left( \theta ,t\right) _{N}$
needs to be evaluated. After simple algebraic calculation, one can get

\begin{equation}
\label{e16}
\partial _{\theta }\left\vert \widehat{\rho}_{s}\left( \theta ,t\right)
\right\rangle \rangle _{N}=\left\{ \widetilde{\Xi} _{\theta }-\frac{1}{2}\partial
_{\theta }\ln \left( \text{Tr}\left[ \widehat{\rho}_{s}\left( \theta ,t\right)
^{2}\right] \right) \right\} \left\vert \widehat{\rho}_{s}\left( \theta
,t\right) \right\rangle \rangle _{N}.
\end{equation}%

Finally, substituting Eqs.~(\ref{e15})-(\ref{e16}) into Eq.~(\ref{e14}) one can obtain
\begin{widetext}
	\begin{eqnarray}
	\label{e17}
	\widetilde{F}\left( \theta ,t\right)  &=&4\left(_{N}\left\langle \langle
	\partial _{\theta }\widehat{\rho}_{s}\left( \theta ,t\right) \right\vert
	\partial _{\theta }\widehat{\rho}_{s}\left( \theta ,t\right) \rangle \rangle
	_{N}-\left\vert _{N}\left\langle \langle \widehat{\rho}_{s}\left( \theta
	,t\right) \right\vert \partial _{\theta }\widehat{\rho}_{s}\left( \theta
	,t\right) \rangle \rangle _{N}\right\vert ^{2}\right)   \notag \\
	&=&4\left( _{N}\left\langle \langle \widehat{\rho}_{s}\left( \theta ,t\right)
	\right\vert \widetilde{\Xi} _{\theta }^{\dag }\widetilde{\Xi} _{\theta }\left\vert \widehat{\rho}%
	_{s}\left( \theta ,t\right) \right\rangle \rangle _{N}-\text{ }_{N}\langle
	\left\langle \widehat{\rho}_{s}\left( \theta ,t\right) \right\vert \widetilde{\Xi} _{\theta
	}^{\dag }\left\vert \widehat{\rho}_{s}\left( \theta ,t\right) \right\rangle
	\rangle _{N}\left\langle \langle \widehat{\rho}_{s}\left( \theta ,t\right)
	\right\vert \widetilde{\Xi} _{\theta }\left\vert \widehat{\rho}_{s}\left( \theta ,t\right)
	\right\rangle \rangle _{N}\right)   \notag \\
	&=&4\text{Cov}_{\left\vert \widehat{\rho}_{s}\left( \theta,t\right) \rangle
		\right\rangle _{N}}\left( \widetilde{\Xi} _{\theta }^{\dag },\widetilde{\Xi} _{\theta }\right) ,
	\end{eqnarray}%
\end{widetext}
where Cov$_{\left\vert \widehat{\rho}_{s}\rangle \right\rangle }\left(
A,B\right) =\left\langle AB\right\rangle _{\left\vert \widehat{\rho}_{s}\rangle
	\right\rangle }-\left\langle A\right\rangle _{\left\vert \widehat{\rho}%
	_{s}\rangle \right\rangle }\left\langle B\right\rangle _{\left\vert \widehat{\rho%
	}_{s}\rangle \right\rangle }$, representing the covariance between variables $A$ and $B$
under state $\left\vert \widehat{\rho}_{s}\rangle \right\rangle$. It is easy to verify $\widetilde{F}\left( \theta,t\right)$ is a real number. This is the second core formula in the current work. 

One can then construct the dissipative quantum Cram\'er-Rao bound (DQCRB) to quantify estimation precision in Liouville space by analogy with the conventional parameter estimation framework, that is, 
\begin{equation}
\label{e23f}
\text{Var}( \widehat{\theta}) \geq \frac{1}{n\widetilde{F}\left( \theta
	,t\right) },
\end{equation}
where $n$ is the number of metrology protocols.
Further, the ``uncertainty relation'' 
\begin{equation}
\langle \Delta \widehat{\theta}%
^{2}\rangle \text{Cov}\left( \widetilde{\Xi} _{\theta }^{\dag },\widetilde{\Xi} _{\theta }\right) \geq 1/4
\end{equation}
is constructed for a single metrology protocol. 
For a closed system, $\widetilde{\Xi} _{\theta }$ degenerates into a Hermitian operator, the above formula is transformed into the form $\langle \Delta \widehat{\theta}^{2}\rangle \langle \Delta \widetilde{\Xi} _{\theta }^{2}\rangle\geq 1/4$, which is completely consistent with the form of Eq.~(\ref{e6aa}). 
Moreover, for this special case, the condition for DQFI to get its maximum value and the expression for the corresponding maximum value are similar to Eqs.~(\ref{e7aa}) and (\ref{e6}), respectively, as long as the conventional generator $\widehat{h}_{\theta }$ is replaced with $\widetilde{\Xi} _{\theta }$. 
However, one should be noted that for pure states (in lossless case), the maximum values of DQFI are twice that of CQFI, i.e., $\widetilde{F}^{\text{max}}=2F_{q}^{\text{max}}$, because there are more potential metrology resources in the extended state space (Liouville space). The proof can be found in Appendix \ref{AC} [see Eq.~(\ref{e106})].

For general scenarios, the DQCRB conveys an intuition that the greater correlation between $\widetilde{\Xi} _{\theta }^{\dag }$ and $\widetilde{\Xi} _{\theta }$, the higher estimation precision of unknown parameters. In addition, although Eqs.~(\ref{e5}) and (\ref{e17}) are very similar in form, they are essentially different because the CQFI is proportional to the self variance of the conventional generator $\widehat{h}_{\theta }$. 
This difference stems from the fact that the generator in the closed system is Hermitian, but not in the open system. More importantly, Eq.~(\ref{e17})  directly links the ultimate precision limit with the general underlying dynamics of the dissipative system in the Liouville space. This is difficult to achieve under the CQFI framework owing to the abstract superoperator.

Particularly, supposing $\theta$ is an overall factor, i.e., $\vec{L}(\theta ,t)= \theta \left(
t\right) \vec{L}$ and using the derivative chain rule $\partial _{\theta }\widetilde{U}_{\theta }=( \partial _{t}\widetilde{U}_{\theta }) \left( \partial t/\partial \theta \right) $, Eq.~(\ref{e17}) can be simplified as
\begin{equation}
\widetilde{F}\left( \theta ,t\right)  =\frac{4}{\left[ \partial _{t}\ln \theta
	\left( t\right) \right] ^{2}}\text{Cov}_{\left\vert \widehat{\rho}_{s}\left(
	\theta ,t\right) \rangle \right\rangle _{N}}\left( \vec{L}^{\dag },\vec{L}%
\right) , \label{e18}
\end{equation}%
which is consistent with Eq.~(5) of  Ref. \cite{alipour2014quantum}. This justifies the universality of our formula. Additionally, if $\theta$ does not change with time, one can further obtain  
\begin{equation}
\widetilde{F}\left( \theta ,t\right)  =4t^2\text{Cov}_{\left\vert \widehat{\rho}_{s}\left(
	\theta ,t\right) \rangle \right\rangle _{N}}\left( \vec{L}^{\dag },\vec{L}%
\right). \label{e18a}
\end{equation}%
In contrast to the CQFI in Eq.~(\ref{e7}), although the DQFI also has a time-squared factor in this case, the covariance is time-dependent and therefore does not reach the Heisenberg precision unless there is no dissipation and Hamiltonian $\widehat{H}$ is time-independent in which case $\vec{L}$ is only a replicative extension of $\widehat{H}$. In the presence of dissipation, the scaling of estimation error given by the DQFI is approximately Heisenberg-scale-like in a very short time. As time increases, however, the variation of scaling of estimation error cannot be judged due to the complexity of open system dynamics, which strongly relies on the feature and intensity of the noise suffered. We will illustrate this with a specific system in Sec.~\ref{IV}.

According to Eq.~(\ref{e17}), in order to maximize $\widetilde{F}\left( \theta ,t\right)$, it is necessary to maximize the covariance between $\widetilde{\Xi} _{\theta }^{\dag }$ and $\widetilde{\Xi} _{\theta }$ respect to the  coherence vector $\left\vert \widehat{\rho}_{s}\left( \theta ,t\right) \rangle \right\rangle _{N}$. 
Unfortunately, due to the mathematical difficulties, one can't directly obtain, as in a closed system, the maximum value of $\widetilde{F}\left( \theta,t\right)$ as well as the condition under which  $\widetilde{F}\left( \theta,t\right)$ gets the maximum value. But if we define two Hermitian generator operators in the Liouville space, i.e.,
\begin{equation}
\label{e20a}
\widetilde{\Theta}_{\theta }=\frac{\widetilde{\Xi}_{\theta }+\widetilde{\Xi}_{\theta
	}^{\dag }}{2},\widetilde{\Lambda}_{\theta }=\frac{i(\widetilde{\Xi}_{\theta }-\widetilde{%
		\Xi}_{\theta }^{\dag })}{2},
\end{equation}%
and substitute them into Eq.~(\ref{e17}), then using the Heisenberg uncertainty relation $\langle \Delta \widetilde{\Theta}_{\theta }^{2}\rangle \langle \Delta \widetilde{%
\Lambda}_{\theta }^{2}\rangle \geq |i[\widetilde{\Lambda}_{\theta },\widetilde{\Theta%
}_{\theta }]|^{2}/4$ \cite{ballentine2014quantum} and the basic inequality $\left\langle
\Delta \aleph ^{2}\right\rangle \leq \left( \lambda _{\max }^{\aleph
}-\lambda _{\min }^{\aleph }\right) ^{2}/4$ \cite{giovannetti2006quantum} where $\lambda _{\max }^{\aleph }$ and $\lambda _{\min }^{\aleph }$ refer to the maximum and minimum eigenvalues of operator $\aleph\in\{\widetilde{\Theta}_{\theta },\widetilde{\Lambda}_{\theta }\}$, respectively, can help us obtain an inequality of DQFI,
\begin{widetext}
	\begin{equation}
	\widetilde{F}\left( \theta ,t\right)   =4\left[ \left\langle \Delta \widetilde{\Theta}_{\theta }^{2}\right\rangle
	_{\left\vert \widehat{\rho}_{s}\left( \theta ,t\right) \rangle \right\rangle
		_{N}}+\left\langle \Delta \widetilde{\Lambda}_{\theta }^{2}\right\rangle
	_{\left\vert \widehat{\rho}_{s}\left( \theta ,t\right) \rangle \right\rangle
		_{N}}+ \left\langle i[\widetilde{\Lambda}_{\theta },\widetilde{\Theta}_{\theta
	}]\right\rangle _{\left\vert \widehat{\rho}_{s}\left( \theta, t\right)
		\rangle \right\rangle _{N}}\right]   
	\leq \left[ \left( \lambda _{\max }^{\widetilde{\Theta}_{\theta }}+\lambda
	_{\max }^{\widetilde{\Lambda}_{\theta }}\right) -\left( \lambda _{\min }^{\widetilde{%
			\Theta}_{\theta }}+\lambda _{\min }^{\widetilde{\Lambda}_{\theta }}\right) %
	\right] ^{2}.  \label{e20}
	\end{equation}
\end{widetext}
This inequality indicates that the upper bound of DQFI is determined by the maximum (minimum) eigenvalues of generators $\widetilde{\Theta}_{\theta }$ and $\widetilde{\Lambda}_{\theta }$, namely the highest estimation precision under the extended description is limited by the underlying dynamics of the open system, which is very similar to that of the closed system [see Eq.~(\ref{e6})]. 
It should be noted that this bound does not necessarily give the tightest upper bound, as an optimal state $\left\vert \widehat{\rho}_{s}\left( \theta,t\right) \rangle \right\rangle_{N}$ in which ``$\leq$'' in Eq.~(\ref{e20}) takes the equal sign may not be found. 
More importantly, we should recognize that Eq.~(\ref{e20a}) is not just a mathematical transformation on the dissipative generator $\widetilde{\Xi} _{\theta }$ to estimate the upper bound, but has a clear physical meaning, that is, it transforms a non-unitary encoding process into two unitary encoding processes. In exchange, a  part of DQFI comes from their noncommutability, namely the commutator of $\widetilde{\Theta}_{\theta }$ and $\widetilde{\Lambda}_{\theta }$.

Next, we turn the target to the calculation of $\widetilde{\Xi} _{\theta }$ and $\widetilde{\Xi} _{\theta
}^{\dag }$ in the Liouville space. Here, for simplicity, we focus on the case that the Liouvillian superoperator $\widehat{\mathcal{L}}(\theta,t)$ is independent of time, so supermatrix $\vec{L}%
(\theta ,t)$ $\rightarrow \vec{L}(\theta )$ and $\widetilde{U}_{\theta }=e^{\vec{%
		L}(\theta )t}$. Generally speaking, $[ \vec{L}(\theta ),\partial _{\theta }\vec{%
	L}(\theta )] =0$ does not hold, which is an important difference from 
Ref. \cite{alipour2014quantum}. For this purpose, we need to utilize a famous
exponential operator integration formula \cite{wilcox1967exponential}
\begin{equation}
\frac{\partial e^{G}}{\partial \lambda }=\int_{0}^{1}e^{sG}\tfrac{\partial G%
}{\partial \lambda }e^{(1-s)G}ds,
\end{equation}%
where $G$ refers to an operator. With the above formula, one can get
\begin{eqnarray}
\frac{\partial \widetilde{U}_{\theta }}{\partial \theta } &=&\int_{0}^{1}e^{s%
	\vec{L}(\theta )t}\tfrac{t\partial \vec{L}(\theta )}{\partial \theta }%
e^{(1-s)\vec{L}(\theta )t}ds  \notag \\
&=&\int_{0}^{t}e^{\mu \vec{L}(\theta )}\tfrac{\partial \vec{L}(\theta )}{%
	\partial \theta }e^{\left( t-\mu \right) \vec{L}(\theta)}d\mu,
\end{eqnarray}%
where $\mu=ts$. Then, based on the definition of $\widetilde{\Xi}_{\theta }$ in Eq. (\ref{e12}), one has
\begin{eqnarray}
\label{e23}
\widetilde{\Xi}_{\theta } &=&i \int_{0}^{t}e^{\mu \vec{L}(\theta )}\tfrac{\partial 
	\vec{L}(\theta )}{\partial \theta }e^{-\mu \vec{L}(\theta )}d\mu , \\ 
\widetilde{\Xi}_{\theta }^{\dag } &=-i &\int_{0}^{t}e^{-\mu \vec{L}^{\dag }(\theta )}%
\tfrac{\partial \vec{L}^{\dag }(\theta )}{\partial \theta }e^{\mu \vec{L}%
	^{\dag }(\theta )}d\mu . \label{e24}
\end{eqnarray}%
In the short-time limit $t\ll 1,\widetilde{\Xi}_{\theta }$ and $\widetilde{\Xi}%
_{\theta }^{\dag }$ can be approximately written as   
\begin{equation}
\widetilde{\Xi}_{\theta }\simeq it\dfrac{\partial \vec{L}(\theta )}{\partial
	\theta },\widetilde{\Xi}_{\theta }^{\dag }\simeq -it\dfrac{\partial \vec{L}^{\dag
	}(\theta )}{\partial \theta }.
\end{equation}
It therefore can be predicted that when the time is very short, DQFI is approximately proportional to $t^2$, which is consistent with our previous analysis below Eq.~(\ref{e18a}).

In order to find the specific forms of $\widetilde{\Xi}_{\theta }$ and $\widetilde{\Xi%
}_{\theta }^{\dag }$ in the whole time domain, we define operator 
\begin{equation}
\Re \left( \mu \right) =-i\frac{\partial \widetilde{\Xi}_{\theta }}{\partial t}|_{t=\mu }=e^{\mu \vec{L}(\theta )}\dfrac{\partial \vec{L}%
	(\theta )}{\partial \theta }e^{-\mu \vec{L}(\theta )},
\end{equation}%
with $\Re \left( 0\right) =\partial _{\theta }\vec{L}(\theta )$. The derivative of $\Re \left( \mu \right) $ with respect to 
$\mu $ satisfies the following relationship,
\begin{eqnarray}
\frac{d\Re \left( \mu \right) }{d\mu } &=&\vec{L}(\theta )\Re \left( \mu
\right) -\Re \left( \mu \right) \vec{L}(\theta)  \notag \\
&=&[\vec{L}(\theta),\Re \left( \mu \right) ]. \label{e27}
\end{eqnarray}

In addition, imitating the Liovillian superoperator $\widehat{\mathcal{L}}$ introduced in Hilbert space which admits both left and right eigenmatrices \cite{minganti2019quantum}, we introduce the following superoperators in the Liouville space, 
\begin{equation}
\mathbf{\widehat{\Pi}}_{\theta }\left[ \bullet \right] =\left[ \vec{L}(\theta
),\bullet \right] ,\mathbf{\widehat{\Pi}}_{\theta }^{\dag }\left[ \bullet \right] =%
\left[ \vec{L}^{\dag }(\theta ),\bullet \right] .
\end{equation}%
In general, $\mathbf{\widehat{\Pi}}_{\theta }$ and $\mathbf{\widehat{\Pi}}_{\theta }^\dagger$ are non-Hermitian superoperator, which satisfying the following eigenequations 
\begin{eqnarray}
\mathbf{\widehat{\Pi}}_{\theta }\widetilde{\Upsilon}_{k} &=&\lambda _{k}\widetilde{\Upsilon}_{k}, \label{eigenPi}\\
\mathbf{\widehat{\Pi}}_{\theta }^{\dag } \widetilde{\Omega}_{k} &=&\lambda _{k}^{\ast
}\widetilde{\Omega}_{k},\label{eigenPi2}
\end{eqnarray}
where $\lambda _{k}$ $( \lambda _{k}^{\ast }) $ and $\widetilde{\Upsilon}_{k}$ $( \widetilde{\Omega}_{k}) $ refer to the eigenvalues and corresponding right (left) eigenmatrices of $\mathbf{\widehat{\Pi}}_{\theta}$, respectively. Here, we do not consider the case of $\widetilde{\Upsilon}_{k}$ ($\widetilde{\Omega}_{k}$) coalesce (namely all $\widetilde{\Upsilon}_{k}$ ($\widetilde{\Omega}_{k}$) are independent), such a situation is
believed to be rare \cite{mori2020resolving}. Note also that the degeneracy of $\lambda _{k}$ does not necessarily
lead to the coalesce of $\widetilde{\Upsilon}_{k}$ ($\widetilde{\Omega}_{k}$). We will see this later.  One can normalize the right eigenmatrices $||\widetilde{\Upsilon}_{k}||^{2}=$Tr$[%
\widetilde{\Upsilon}_{k}^{^{\dag }}\widetilde{\Upsilon}_{k}]=1$ by using the
Hilbert-Schmidt inner product \cite{chen2019sensitivity,minganti2019quantum}%
, and different eigenmatrices may be not orthogonal due to the non-Hermitian
property of $\mathbf{\widehat{\Pi}}_{\theta },$ namely Tr$[\widetilde{\Upsilon}%
_{m}^{^{\dag }}\widetilde{\Upsilon}_{n}]\neq \delta _{mn}$ and Tr$[\widetilde{\Omega}%
_{m}^{^{\dag }}\widetilde{\Omega}_{n}]\neq \delta _{mn}$, in which $\delta _{mn}$
is the Kronecker delta function. Importantly, $||\widetilde{\Omega}_{k}||^{2}=$ Tr%
$[\widetilde{\Omega}_{k}^{^{\dag }}\widetilde{\Omega}_{k}]\neq 1$ is determined by
the mutually orthonormal condition of the right and left eigenmatrices that
Tr$[\widetilde{\Omega}_{m}^{\dag }\widetilde{\Upsilon}_{n}]=\delta _{mn}$ \cite%
{chen2019sensitivity}. In particular, although different right eigenmatrices 
$\widetilde{\Upsilon}_{k}$ are not orthogonal to each other, they can still form
a complete basis $\{\widetilde{\Upsilon}_{k}\}$ \cite%
{minganti2019quantum,mori2020resolving,macieszczak2016towards}. Thus one can
expand the operator $\Re\left( \mu \right) $ as follows
\begin{equation}
\Re \left( \mu \right) =\sum_{k=1}^{d}c_{k}\left( \mu \right) \widetilde{\Upsilon%
}_{k},  \label{e31}
\end{equation}%
where $c_{k}\left( \mu \right) =$ Tr$[ \widetilde{\Omega}_{k}^{^{\dag
}}\partial _{\theta }\vec{L}(\theta )] e^{\lambda _{k}\mu }$, and $%
\Re \left( 0\right) =\partial _{\theta }\vec{L}(\theta )$ has been utilized, and $d$ is the number of $\lambda _{k}$ (perhaps there are eigenvalues degeneracy). It is easy to verify that Eq.~(\ref{e31}) is the solution of Eq.~(\ref{e27}).  
Without losing generality, we suppose that $\lambda _{k}=0$ $(k=1,2,...,r)$, other $\lambda _{k}\neq 0$ $\left( k=r+1,r+2,...,d\right),$ hence Eq.~(\ref{e31}) can be rewritten as 
	\begin{eqnarray}
	\Re \left( \mu \right)  &=&\sum_{k=1}^{r}\text{Tr}\left[ \widetilde{\Omega}%
	_{k}^{^{\dag }}\partial _{\theta }\vec{L}(\theta )\right] \widetilde{\Upsilon}%
	_{k}+  \notag  \label{e30} \\
	&&\sum_{k=r+1}^{d}\text{Tr}\left[ \widetilde{\Omega}_{k}^{^{\dag }}\partial
	_{\theta }\vec{L}(\theta )\right] e^{\lambda _{k}\mu }\widetilde{\Upsilon}_{k}.
	\end{eqnarray}%
	Substituting Eqs.~(\ref{e30}) into Eqs.~(\ref{e23})-(\ref{e24}), we obtain,
	\begin{eqnarray}
	\widetilde{\Xi}_{\theta } &=&it\sum_{k=1}^{r}\text{Tr}\left[ \widetilde{\Omega}%
	_{k}^{^{\dag }}\partial _{\theta }\vec{L}(\theta )\right] \widetilde{\Upsilon}%
	_{k}+  \notag  \label{e33} \\
	&&\sum_{k=r+1}^{d}\frac{i\left( e^{\lambda _{k}t}-1\right) }{\lambda _{k}}%
	\text{Tr}\left[ \widetilde{\Omega}_{k}^{^{\dag }}\partial _{\theta }\vec{L}%
	(\theta )\right] \widetilde{\Upsilon}_{k}, \\
	\widetilde{\Xi}_{\theta }^{\dag } &=&-it\sum_{k=1}^{r}\text{Tr}\left[ \widetilde{%
		\Omega}_{k}\partial _{\theta }\vec{L}^{\dag }(\theta )\right] \widetilde{\Upsilon%
	}_{k}^{^{\dag }}-  \notag \\
	&&\sum_{k=r+1}^{d}\frac{i\left( e^{\lambda _{k}^{\ast }t}-1\right) }{\lambda
		_{k}^{\ast }}\text{Tr}\left[ \widetilde{\Omega}_{k}\partial _{\theta }\vec{L}%
	^{\dag }(\theta )\right] \widetilde{\Upsilon}_{k}^{^{\dag }}.  \label{e34}
	\end{eqnarray}%
Here, the first term is the contribution of $\lambda _{k}=0$ $(k=1,2,...r)$, while the second term originates from $\lambda _{k}\neq 0$ $\left(
	k=r+1,r+2,...,d\right) $. Currently, the forms of $\widetilde{\Xi}_{\theta }$ and $\widetilde{\Xi}_{\theta }^{\dag }$  is still not conducive to our analysis of the origin of DQFI. In particular, if one knows the eigenvalues and the right (left) eigenmatrices of $\mathbf{\widehat{\Pi}}_{\theta }$, Eqs.~(\ref{e33}) and (\ref{e34}) can be further simplified. Later we will see that the reduced form is very useful for us to track the imprint of the estimated parameter, i.e., how the parameter to be estimated is encoded into the open system. Next, we start performing the simplification of $\widetilde{\Xi}_{\theta }$ and $\widetilde{\Xi}_{\theta }^{\dag }$.
 
Since the Liouvillian supermatrix $\vec{L}(\theta )$ is also non-Hermitian, its left and right eigenvectors need to be considered. The eigenequations of $\vec{L}(\theta )$ satisfies the following relationship%
\begin{eqnarray}
\vec{L}(\theta )\left\vert \phi _{n}\right\rangle \rangle 
&=&L_{n}\left\vert \phi _{n}\right\rangle \rangle , \\
\vec{L}^{\dag }(\theta )\left\vert \chi _{n}\right\rangle \rangle 
&=&L_{n}^{\ast }\left\vert \chi _{n}\right\rangle \rangle ,
\end{eqnarray}%
where $L_{n}$ $\left( L_{n}^{\ast }\right) $\ and $\left\vert \phi
_{n}\right\rangle \rangle $ $\left( \left\vert \chi _{n}\right\rangle
\rangle \right) $ denote the eigenvalues and corresponding right
(left) eigenvectors of $\vec{L}(\theta )$, respectively. We first do not consider the case of $L_{n}$ degeneracy. The eigenvalues can be sorted as $%
0=L_{1}>$ Re$(L_{2})$ $\geq $ Re$(L_{3})\ldots \geq$ Re $(L_{M^{2}})$ because Liouvillian supermatrix $\vec{L}(\theta )$ is negative semidefinite, here $M$ being the original Hilbert-space dimension.  This is an inexorable requirement for the open system to be stable in the long-term limit, otherwise, the system tends to diverge [if Re$(L_{k})>0$ exists].
Similar to the case of eigenmatrices $\widetilde{\Upsilon}_{k}$ and $\widetilde{\Omega}_{k}$ of the superoperator $\mathbf{\widehat{\Pi}}_{\theta }$, the Hilbert-Schmidt inner
product of the eigenvectors satisfies $\langle \left\langle \phi _{n}|\phi
_{m}\right\rangle \rangle \neq \delta _{mn}$ and $\langle \left\langle \chi
_{n}|\chi _{m}\right\rangle \rangle \neq \delta _{mn}$, that is, different eigenvectors do not satisfy orthogonality with each other \cite{brody2013biorthogonal,curtright2007biorthogonal,chang2013biorthogonal}. 
One assumed that the right eigenvectors satisfies normalization $\langle \left\langle \phi _{n}|\phi
_{n}\right\rangle \rangle =1$, while the corresponding left eigenvector
satisfies $\langle \left\langle \chi _{n}|\chi _{n}\right\rangle \rangle
\geq 1$ \cite{brody2013biorthogonal} and is determined by the conditions that $\langle \left\langle \chi
_{n}|\phi _{m}\right\rangle \rangle =\delta _{mn}$. Therefore, left and right
eigenvectors satisfy orthogonality. Notice also that $\left\{ \left\vert \phi
_{n}\right\rangle \rangle \right\} $ and $\left\{ \left\vert  \chi
_{n}\right\rangle \rangle \right\} $ can form a complete biorthogonal basis
\cite{brody2013biorthogonal,curtright2007biorthogonal,chang2013biorthogonal}, i.e., $\sum_{n=1}^{M^{2}}\left\vert
\phi _{n}\right\rangle \rangle \langle \left\langle \chi _{n}\right\vert
=\sum_{n=1}^{M^{2}}\left\vert \chi _{n}\right\rangle \rangle \langle
\left\langle \phi _{n}\right\vert =\mathbb{1}_{M^2}$.

Interestingly, we find  that the eigenmatrices of superoperator $\mathbf{\widehat{\Pi}}_{\theta }$ can be built by the eigenvectors of operator $\vec{L}(\theta )$, that is, $\widetilde{\Upsilon}_{k}^{n,m}=\left\vert
\phi _{n}\right\rangle \rangle \langle \left\langle \chi _{m}\right\vert $
and $\widetilde{\Omega}_{k}^{n,m}=\left\vert \chi _{n}\right\rangle \rangle
\langle \left\langle \phi _{m}\right\vert $ so that the eigenequations~(\ref{eigenPi}) and (\ref{eigenPi2}) becomes
\begin{eqnarray}
\label{e37}
\mathbf{\widehat{\Pi}}_{\theta }\widetilde{\Upsilon}_{k}^{n,m} &=&\left[ \vec{L}%
(\theta ),\widetilde{\Upsilon}_{k}^{n,m}\right] =\left( L_{n}-L_{m}\right) 
\widetilde{\Upsilon}_{k}^{n,m}, \\
\mathbf{\widehat{\Pi}}_{\theta }^{\dag}\widetilde{\Omega}_{k}^{n,m} &=&\left[ \vec{L}^{\dag}(\theta ),\widetilde{\Omega}_{k}^{n,m}\right] =\left( L_{n}^{\ast }-L_{m}^{\ast
}\right) \widetilde{\Omega}_{k}^{n,m}, \label{e38}
\end{eqnarray}%
where the eigenvalues corresponding to eigenmatrices $\widetilde{\Upsilon}_{k}^{n,m}$ and $\widetilde{\Omega}_{k}^{n,m}$ are $\left( L_{n}-L_{m}\right) $ and $\left( L_{n}^{\ast
}-L_{m}^{\ast }\right) $, respectively. 
Particularly, in the case of $n=m$, i.e., the eigenvalue of $\mathbf{\widehat{\Pi}}_{\theta }$ is 0, one can obtain
\begin{eqnarray}
\label{e39}
\text{Tr}\left[ \widetilde{\Omega}_{k}^{n,n^{\dag }}\partial _{\theta }\vec{L}%
(\theta )\right]  &=&\sum_{i=1}^{M^{2}}\langle \left\langle \chi
_{i}\right\vert \left\vert \phi _{n}\right\rangle \rangle \langle
\left\langle \chi _{n}\right\vert \partial _{\theta }\vec{L}(\theta
)\left\vert \phi _{i}\right\rangle \rangle   \nonumber \\
&=&\langle \left\langle \chi _{n}\right\vert \partial _{\theta }\vec{L}%
(\theta )\left\vert \phi _{n}\right\rangle \rangle   \nonumber \\
&=&\partial _{\theta }L_{n},
\end{eqnarray}%
where we have used the completeness of biorthogonal basis and $\partial _{\theta }\langle \left\langle \chi
_{n}\right\vert \phi _{n}\rangle \rangle =0$. In this case, one can see that the eigenvalues corresponding to $\widetilde{\Upsilon}_{k}^{n,n}$ $(n=1,2,\ldots, M^{2})$ are all $0$, but $\widetilde{\Upsilon}_{k}^{n,n}$ are independent of each other for different $n$.
By comparison, in the case of $n\neq m$, i.e., the eigenvalue of $\mathbf{\widehat{\Pi}}_{\theta }$ is not 0, one can get
\begin{eqnarray}
\label{e40}
\text{Tr}\left[ \widetilde{\Omega}_{k}^{n,m^{\dag }}\partial _{\theta }\vec{L}%
(\theta )\right]  &=&\sum_{i=1}^{M^{2}}\langle \left\langle \chi
_{i}\right\vert \left\vert \phi _{m}\right\rangle \rangle \langle
\left\langle \chi _{n}\right\vert \partial _{\theta }\vec{L}(\theta
)\left\vert \phi _{i}\right\rangle \rangle   \nonumber \\
&=&\langle \left\langle \chi _{n}\right\vert \partial _{\theta }\vec{L}%
(\theta )\left\vert \phi _{m}\right\rangle \rangle \nonumber  \\
&=&\left( L_{n}-L_{m}\right) \langle \langle \partial _{\theta }\chi
_{n}\left\vert \phi _{m}\right\rangle \rangle ,
	\end{eqnarray}
in which $\langle \left\langle \partial _{\theta}\chi _{n}\right\vert \phi
_{m}\rangle \rangle =-\langle \left\langle \chi _{n}\right\vert \partial
_{\theta}\phi _{m}\rangle \rangle $ is used. Substituting Eqs.~(\ref{e39})-(\ref{e40}) into Eqs.~(\ref{e33}) and (\ref{e34}), we  finally obtain
\begin{widetext}
	\begin{eqnarray}
	\label{e41}
	\widetilde{\Xi}_{\theta } &=&it\sum_{n=1}^{M^{2}}\partial _{\theta}L_{n}\left\vert
	\phi _{n}\right\rangle \rangle \langle \left\langle \chi _{n}\right\vert
	+\sum_{n,m=1,n\neq m}^{M^{2}}i\left[ e^{\left( L_{n}-L_{m}\right) t}-1\right]
	\langle \langle \partial _{\theta }\chi _{n}\left\vert \phi
	_{m}\right\rangle \rangle \left\vert \phi _{n}\right\rangle \rangle \langle
	\left\langle \chi _{m}\right\vert , \\
	\widetilde{\Xi}_{\theta }^{\dag } &=&-it\sum_{n=1}^{M^{2}}\partial _{\theta}L_{n}^{\ast
	}\left\vert \chi _{n}\right\rangle \rangle \langle \left\langle \phi
	_{n}\right\vert -\sum_{n,m=1,n\neq m}^{M^{2}}i\left[ e^{\left( L_{n}^{\ast
		}-L_{m}^{\ast }\right) t}-1\right] \langle \langle \phi _{m}\left\vert
	\partial _{\theta }\chi _{n}\right\rangle \rangle \left\vert \chi
	_{m}\right\rangle \rangle \langle \left\langle \phi _{n}\right\vert . \label{e42}
	\end{eqnarray}
\end{widetext}
Particularly, one can see that at this point the expressions of $\widetilde{\Xi}_{\theta }$ and $\widetilde{\Xi}_{\theta }^{\dag }$ consist entirely of the eigenvalues and eigenvectors of the Liouvillian supermatrix $\vec{L}(\theta )$ that determine the dynamics of the open system. This is the third core formula of current work, and their forms contain rich implications in physics. 

Considering that DQFI is proportional to the covariance between $\widetilde{\Xi}_{\theta }$ and $\widetilde{\Xi}_{\theta }^{\dag }$, see Eq.~(\ref{e17}). Eqs.~(\ref{e41}) and (\ref{e42}) indicate that the DQFI originates from the following two parts. Part I: the dependence of the eigenvalues (real or complex numbers) on $\theta $, and this part is proportional to the time $t$. Part II: the change of eigenvectors with $\theta $, and the relation between this part and time $t$ is closely related to the gap in the spectrum of the corresponding Liouville.
Specifically, it can be divided into the following three situations: (i) in the case that both $L_{n}$ and $L_{m}$ are real numbers, part II exponentially gains over time $t$ when $L_{n}>L_{m}$ and exponentially decays when $L_{n}<L_{m}$. 
(ii) in the case that both $L_{n}$ and $L_{m}$ are pure imaginary numbers, part II  harmonically oscillates with time $t$, which is similar to the closed system. (iii) in the case that at least one of $L_{n}$ and $L_{m}$ is a complex number but not a pure imaginary number, part II shows exponentially-gained oscillations over time $t$ when $\text{Re}\left( L_{n}\right) >\text{Re}\left( L_{m}\right) $ and exponentially-decayed oscillations when $\text{Re}\left( L_{n}\right) <\text{Re}\left( L_{m}\right)$. 
By comparison, from the decomposed form of generator $\widehat{h}_{\theta }$, see Eq.~(28) of Ref. \cite{pang2014quantum}, only the dependence of the real eigenvalues on $\theta $ and harmonic oscillations with time are presented. 

Therefore, in the DQFI framework based on the Liouville space, we have well characterized the possible impact of external environment on estimation precision through the complex eigenvalues and eigenvectors of Liouvillian supermatrix $\vec{L}(\theta)$. It is very difficult to attain this in the CQFI framework based on the Hilbert space owing to the abstract superoperator. In particular, we point out that although the evolution of DQFI with time is closely related to the dependence of $\widetilde{\Xi}_{\theta }$ and $\widetilde{\Xi}_{\theta }^{\dag }$ on time, the final DQFI is determined by the covariance of $\widetilde{\Xi}_{\theta }$ and $\widetilde{\Xi}_{\theta }^{\dag }$ concerning the output state. Accordingly, even if the dissipative generator contains the term of exponential gain over time, it may be offset by the exponential decay factor contained in the output state. As a result, DQFI does not exhibit the non-physical phenomenon of exponential divergence over time under the influence of noise. 

Next, we analyze the DQFI in the long-time limit with the help of Eqs.~(\ref{e41}) and (\ref{e42}).
It is well known that the intrinsically irreversible dynamics of an open system leads the system to the ultimate steady state in a unidirectional time direction. Here, according to the Sch\"{o}dinger-liked equation~(\ref{e9}) the steady state of the open system in the
Liuville space is $\left\vert \widehat{\rho}_{s}\left( g,t\rightarrow +\infty
\right) \rangle \right\rangle _{N}=\left\vert \phi _{1}\right\rangle \rangle$ whose eigenvalue $L_{1}=0$ implying the slowest decay rate \cite{breuer2002theory,rivas2012open} because the eigenvectors corresponding to eigenvalues with real parts less than zero will be attenuated. To explore the  behaviour of $\widetilde{F}_{\theta }\left( t\right)$ at the long-term limit, we assume that
\begin{subequations}
	\label{e44}
	\begin{eqnarray}
	L_{n} &=&x_{n}+iy_{n},L_{m}=x_{m}+iy_{m}, \\  
	\upsilon _{nm} &=&x_{n}-x_{m},\beta _{nm}=y_{n}-y_{m}, 
	\end{eqnarray}%
\end{subequations}
where $x_{n(m)}$ and $y_{n(m)}$ are real part and imaginary part of eigenvalues $L_{n(m)}$, respectively, and $\upsilon_{nm}$ and $\beta_{nm}$ are their difference. Substituting Eq.~(\ref{e44}) into Eqs.~(\ref{e41}) and (\ref{e42}), $\widetilde{\Xi}_{\theta }$ and $\widetilde{\Xi}_{\theta }^{\dag }$ can be rewritten as
\begin{widetext}
	\begin{eqnarray}
	\label{e45}
	\widetilde{\Xi}_{\theta } &=&t\sum_{n=1}^{M^{2}}\partial _{\theta }\left(
	ix_{n}-y_{n}\right) \left\vert \phi _{n}\right\rangle \rangle \langle
	\left\langle \chi _{n}\right\vert +\sum_{n,m=1,n\neq m}^{M^{2}}i\left[
	e^{\left( \upsilon _{nm}+i\beta _{nm}\right) t}-1\right] \langle \langle
	\partial _{\theta }\chi _{n}\left\vert \phi _{m}\right\rangle \rangle
	\left\vert \phi _{n}\right\rangle \rangle \langle \left\langle \chi
	_{m}\right\vert , \\
	\widetilde{\Xi}_{\theta }^{\dag } &=&t\sum_{n=1}^{M^{2}}\partial _{\theta
	}\left( -ix_{n}-y_{n}\right) \left\vert \chi _{n}\right\rangle \rangle
	\langle \left\langle \phi _{n}\right\vert -\sum_{n,m=1,n\neq m}^{M^{2}}i\left[
	e^{\left( \upsilon _{nm}-i\beta _{nm}\right) t}-1\right] \langle \langle
	\phi _{m}\left\vert \partial _{\theta }\chi _{n}\right\rangle \rangle
	\left\vert \chi _{m}\right\rangle \rangle \langle \left\langle \phi
	_{n}\right\vert . \label{e46}
	\end{eqnarray}%
	Based on Eqs.~(\ref{e17}), (\ref{e45}), and (\ref{e46}), after tedious algebraic calculation, we finally have
	\begin{eqnarray}
	\widetilde{F}_{\theta }\left( t\right)  &=&4\left[ \left\langle \widetilde{\Xi}%
	_{\theta }^{\dag }\widetilde{\Xi}_{\theta }\right\rangle _{\left\vert \phi
		_{1}\right\rangle \rangle }-\left\langle \widetilde{\Xi}_{\theta }^{\dag
	}\right\rangle _{\left\vert \phi _{1}\right\rangle \rangle }\left\langle 
	\widetilde{\Xi}_{\theta }\right\rangle _{\left\vert \phi _{1}\right\rangle
		\rangle }\right]   \nonumber \\
	&=&4\sum_{n,k=2}^{M^{2}}\left[ e^{\left( \upsilon _{n1}-i\beta _{n1}\right)
		t}-1\right] \left[ e^{\left( \upsilon _{k1}+i\beta _{k1}\right) t}-1\right]
	\langle \langle \phi _{1}\left\vert \partial _{\theta }\chi
	_{n}\right\rangle \rangle \langle \langle \partial _{\theta }\chi
	_{k}\left\vert \phi _{1}\right\rangle \rangle \left[ \langle \langle \phi
	_{n}\left\vert \phi _{k}\right\rangle \rangle -\langle \langle \phi
	_{n}\left\vert \phi _{1}\right\rangle \rangle \langle \langle \phi
	_{1}\left\vert \phi _{k}\right\rangle \rangle \right] . \label{e47}
	\end{eqnarray}%
	
	Considering that $\upsilon _{n1},\upsilon _{k1}<0$ due to $L_{1}=0$ and Re$\left[ L_{n>1}\right] <0$, in the limit case that $t\rightarrow +\infty $, Eq.~(\ref{e47}) becomes into
	\begin{equation}
	\widetilde{F}_{\theta }\left( t\rightarrow +\infty \right) \simeq
	4\sum_{n,k=2}^{M^{2}}\langle \langle \phi _{1}\left\vert \partial _{\theta
	}\chi _{n}\right\rangle \rangle \langle \langle \partial _{\theta }\chi
	_{k}\left\vert \phi _{1}\right\rangle \rangle \left[ \langle \langle \phi
	_{n}\left\vert \phi _{k}\right\rangle \rangle -\langle \langle \phi
	_{n}\left\vert \phi _{1}\right\rangle \rangle \langle \langle \phi
	_{1}\left\vert \phi _{k}\right\rangle \rangle \right] , \label{e48}
	\end{equation}%
\end{widetext}
which gives the value of DQFI when the system tends to a steady state, which depends on the Hilbert-Schmidt inner product between the eigenvectors as well as the inner product between the eigenvectors and the derivative of the eigenvectors with respect to $\theta$.
What is more, it is easy to see that DQFI is not necessarily zero at the end under the influence of noise. For example, for a two-level system affected by dephasing noise, when estimating the parameter only embedded in the probability amplitude of its quantum state, DQFI is not eventually towards zero. This is because dephasing noise only destroys the phase information of the quantum state. In contrast, if the system is affected by spin-flip noise and the state eventually becomes a maximally mixed state, one can derive 
$\widetilde{F}_{\theta }\left( t\rightarrow +\infty \right) =0$ from the equations $\partial_{\theta}\vert \widehat{\rho}_{s}\left( g,t\rightarrow +\infty\right) \rangle \rangle _{N}=\partial_{\theta}\left\vert \phi _{1}\right\rangle \rangle=0 $, $\langle\langle \phi _{1}\left\vert \chi _{n}\right\rangle\rangle =0$,
and $\langle
\langle \phi _{1}\left\vert \partial _{\theta }\chi _{n}\right\rangle\rangle =0$ for $(n>1)$, which is within our
expectation.

It is worth mentioning that in some open systems when the characteristic frequency of the system and decay rate meet certain conditions, leading to the possible appearance of LEP \cite{minganti2019quantum,khandelwal2021signatures,chen2022decoherence}, namely at least two eigenvalues are identical and the corresponding eigenvectors coalesce for the supermatrix $\vec{L}(\theta)$. 
In this exotic case, the left and right eigenvectors of supermatrix $\vec{L}(\theta)$ can no longer form a set of complete biorthogonal basis due to the reduction of independent eigenvectors.

The method of constructing complete biorthogonal basis in the presence of LEP was proposed in Ref. \cite{sarandy2005adiabatic}. The specific steps are as follows: (i) Perform Jordan decomposition of $\vec{L}(\theta )$ through a transformation $\mathbcal{S}$, i.e., $\vec{L}_{J}(\theta )=\mathbcal{S}^{-1}\vec{L}(\theta)\mathbcal{S}$, where $\vec{L}_{J}(\theta)$ satisfies the Jordan canonical form \cite{bhatia2013matrix}. 
(ii) In order to construct a complete biorthogonal basis, some new left and right eigenvectors of $\vec{L}_{J}(\theta)$ ($ \text{e.g.,}  |\phi _{n, i}\rangle \rangle _{J}$) are defined. Their selection must ensure the elegant block structure of $\vec{L}_{J}(\theta )$. 
(iii) These new eigenvectors are inversely transformed by $\mathbcal{S}$  $\left(\text{e.g.,} |\phi _{n, i}\rangle \rangle =\mathbcal{S}|\phi_{n, i}\rangle \rangle _{J}\right)$ and then combined with original independent left and right eigenvectors of $\vec{L}(\theta )$ to form a complete biorthogonal basis, see Ref. \cite{sarandy2005adiabatic} for details. Next, following the same steps as in the derivation of Eqs.~(\ref{e41}) and (\ref{e42}), one can get the dissipative generator in the
presence of LEPs, reads
\begin{widetext}
	\begin{eqnarray}
	\label{e49}
	\widetilde{\Xi}_{\theta } &=&it\sum_{n=1}^{M^{2}}\partial _{\theta
	}L_{n}\sum_{k=1}^{\xi _{n}}\left\vert \phi _{n,k}\right\rangle \rangle
	\langle \left\langle \chi _{n,k}\right\vert +\sum_{n\neq
		m}^{M^{2}}\sum_{i=1}^{\xi _{n}}\sum_{j=1}^{\xi _{m}}i\left[ e^{\left(
		L_{n}-L_{m}\right) t}-1\right] \langle \langle \partial _{\theta }\chi
	_{n,i}\left\vert \phi _{m,j}\right\rangle \rangle \left\vert \phi
	_{n,i}\right\rangle \rangle \langle \left\langle \chi _{m,j}\right\vert ,  \\
	\widetilde{\Xi}_{\theta }^{\dag } &=&-it\sum_{n=1}^{M^{2}}\partial _{\theta
	}L_{n}^{\ast }\sum_{k=1}^{\xi _{n}}\left\vert \chi _{n,k}\right\rangle
	\rangle \langle \left\langle \phi _{n,k}\right\vert -\sum_{n\neq
		m}^{M^{2}}\sum_{i=1}^{\xi _{n}}\sum_{j=1}^{\xi _{m}}i\left[ e^{\left(
		L_{n}^{\ast }-L_{m}^{\ast }\right) t}-1\right] \langle \langle \phi
	_{m,j}\left\vert \partial _{\theta }\chi _{n,i}\right\rangle \rangle
	\left\vert \chi _{m,j}\right\rangle \rangle \langle \left\langle \phi
	_{n,i}\right\vert . 	\label{e50}
	\end{eqnarray}
\end{widetext}
Here $\xi _{n}$ ($\xi _{m}$) is the degeneracy of the eigenvalue $L_{n}$ ($L_{m}$), corresponding to $\xi _{n}$ ($\xi _{m}$)-order LEP (indicating that the possibility of multiple LEPs emergence). The definition of other symbols is consistent with Eqs.~(\ref{e41}) and (\ref{e42}). Compared with the case without LEP, the existence of $\vec{L}(\theta)$ degeneracy only adds the weighted sum of products of these basis vectors, but there is no substantial correction to the changes in $\widetilde{\Xi}_{\theta }$ and $\widetilde{\Xi}_{\theta }^{\dag }$ over time.

As a last remark, let us stress that: in order to simplify the calculation and analysis, we transfer the parameter estimation problem of open systems from Hilbert space to Liouville space through the vectorization method for processing, hence avoiding the difficulties caused by the Liouvillian superoperator. 
However, we argue that this mathematical treatment does not erase the essence of the physical process so that the dependence of CQFI and DQFI on time or other system parameters is similar, although there may be some quantitative deviations due to the dimensional expansion. 
In principle, one could also define a dissipative generator in Hilbert space in the form of a superoperator to calculate CQFI, but due to its abstraction, it is difficult to analyze or even control the variation of CQFI by studying its eigenvalues and eigenvectors as in a closed system. 
In the following example, one can see that DQFI and CQFI do have similar curve profiles and features as the dissipation rate or the time of evolution changes, and DQFI is more advantageous in dissipative parameter estimation in the presence of singularities, i.e., LEPs.

\section{Example: Two-level system with spin-flip noise}\label{IV}     
In this section, we apply a concrete model to demonstrate our theory. In order to be more representative, we consider an open two-level system subjected to spin-flip noise, its dynamical evolution satisfies the following master equation
\begin{equation}
\partial _{t}\widehat{\rho}_{s}\left( \omega ,t\right) =\widehat{\mathcal{L}}(\omega
)\left[ \widehat{\rho}_{s}\left( \omega ,t\right) \right], \label{e51} 
\end{equation}%
with
\begin{eqnarray*}
	\widehat{\mathcal{L}}(\omega )\left[ \widehat{\rho}_{s}\left( \omega ,t\right) %
	\right]  &=&-i[ \widehat{H}\left( \omega \right) ,\widehat{\rho}_{s}\left(
	\omega ,t\right) ] +\gamma _{x}\mathcal{D}\left[ \widehat{\sigma}_{x}\right] \widehat{\rho}_{s}\left( \omega ,t\right) , \\
	\mathcal{D}\left[ \widehat{\sigma}_{x}\right] \widehat{\rho}_{s}\left( \omega ,t\right)  &=&%
	\widehat{\sigma}_{x}\widehat{\rho}_{s}\left( \omega ,t\right) \widehat{\sigma}_{x}^{\dag
	}-\widehat{\rho}_{s}\left( \omega ,t\right) ,
\end{eqnarray*}%
where the Hamiltonian $\widehat{H}\left( \omega \right) =\omega \widehat{\sigma}_{z}/2$ with the transition frequency $\omega$ being the parameter to be estimated. The Lindbrad superoperator $\mathcal{D}\left[ \widehat{\sigma}_{x}\right]$ describes spin-flip noise with decay rate $\gamma _{x}$, $\widehat{\sigma}%
_{x,y,z}$ are the Pauli operators, and $\widehat{\sigma}_{\pm }=(\widehat{\sigma}_{x}\pm i\widehat{\sigma}_{y})/2$ denotes the flip-up (flip-down) operator. 
After vectorizing Eq.~(\ref{e51}), one can get a Sch\"{o}dinger-liked equation,
\begin{equation}
\partial _{t}\left\vert \widehat{\rho}_{s}\left( \omega ,t\right) \rangle
\right\rangle =\vec{L}(\omega ,t)\left\vert \widehat{\rho}_{s}\left( \omega
,t\right) \rangle \right\rangle , \label{e52}
\end{equation}%
where
\begin{equation}
\vec{L}(\omega )=\left[ 
\begin{array}{cccc}
-\gamma _{x} & 0 & 0 & \gamma _{x} \\ 
0 & -\gamma _{x}-i\omega  & \gamma _{x} & 0 \\ 
0 & \gamma _{x} & -\gamma _{x}+i\omega  & 0 \\ 
\gamma _{x} & 0 & 0 & -\gamma _{x}%
\end{array}%
\right] 
\end{equation}%
is Liouvillian supermatrix of $\widehat{\mathcal{L}}(\omega )$ under the $\widehat{%
	\sigma}_{z}$ representation. $[ \partial _{\omega }\vec{L}(\omega ),%
\vec{L}(\omega )] \neq 0$ can be easily verified (non-phase-covariant dynamic), this is
an essential difference from the examples in Refs. \cite{alipour2014quantum,benatti2014dissipative}.  Solving the eigenequation of the supermatrix $\vec{L}(\omega )$, we obtain
\begin{subequations}
	\label{e54}
	\begin{eqnarray}
	L_{1} &=&0,L_{2}=-2\gamma _{x},L_{3\left( 4\right) }=-\gamma _{x}\mp \Omega, \\
	\left\vert \phi _{1}\right\rangle \rangle  &=&\frac{1}{\sqrt{2}}\left[
	1,0,0,1\right] ^{\text{T}}, \\
	\left\vert \phi _{2}\right\rangle \rangle  &=&\frac{1}{\sqrt{2}}\left[
	-1,0,0,1\right] ^{\text{T}}, \\
	\left\vert \phi _{3}\right\rangle \rangle  &=&\alpha _{1\omega }\left[
	0,-i\omega -\Omega ,\gamma _{x},0\right] ^{\text{T}}, \\
	\left\vert \phi _{4}\right\rangle \rangle  &=&\alpha _{2\omega }\left[
	0,-i\omega +\Omega ,\gamma _{x},0\right] ^{\text{T}}, \\
	\left\vert \chi _{1}\right\rangle \rangle  &=&\frac{1}{\sqrt{2}}\left[
	1,0,0,1\right] ^{\text{T}}, \\
	\left\vert \chi _{2}\right\rangle \rangle  &=&\frac{1}{\sqrt{2}}\left[
	-1,0,0,1\right] ^{\text{T}}, \\
	\left\vert \chi _{3}\right\rangle \rangle  &=&\beta _{1\omega }\left[
	0,i\omega -\Omega ^{\ast },\gamma _{x},0\right] ^{\text{T}}, \\
	\left\vert \chi _{4}\right\rangle \rangle  &=&\beta _{2\omega }\left[
	0,i\omega +\Omega ^{\ast },\gamma _{x},0\right] ^{\text{T}}.
	\end{eqnarray}%
\end{subequations}
with 
\begin{eqnarray*}
	\alpha _{1\omega } &=&\frac{1}{\sqrt{\left( \Omega +i\omega \right) \left(
			\Omega ^{\ast }-i\omega \right) +\gamma _{x}^{2}}}, \\
	\alpha _{2\omega } &=&\frac{1}{\sqrt{\left( \Omega -i\omega \right) \left(
			\Omega ^{\ast }+i\omega \right) +\gamma _{x}^{2}}}, \\
	\beta _{1\omega } &=&\frac{\sqrt{\left( \Omega +i\omega \right) \left(
			\Omega ^{\ast }-i\omega \right) +\gamma _{x}^{2}}}{\left( \Omega ^{\ast
		}-i\omega \right) ^{2}+\gamma _{x}^{2}}, \\
	\beta _{2\omega } &=&\frac{\sqrt{\left( \Omega -i\omega \right) \left(
			\Omega ^{\ast }+i\omega \right) +\gamma _{x}^{2}}}{\left( \Omega ^{\ast
		}+i\omega \right) ^{2}+\gamma _{x}^{2}},
\end{eqnarray*}%
where $\Omega =\sqrt{\gamma _{x}^{2}-\omega ^{2}}$. $L_{i}$ is the eigenvalue of $\vec{L}(\omega )$ and $\left\vert \phi _{i}\right\rangle
\rangle $ $\left( \left\vert \chi _{i}\right\rangle \rangle \right) $
denotes the right (left) eigenvector. 
Note that in Eq.~(\ref{e54}), the right eigenvectors have been normalized, and the coefficients $\beta _{1\omega }$ and $\beta _{2\omega }$ in front of the left eigenvectors are to satisfy condition $\langle \left\langle \chi _{n}|\phi _{m}\right\rangle
\rangle =\delta _{mn}$. 
Particularly, one can see that $L_{3}=L_{4}$ and $\left\vert \phi_{3}\right\rangle \rangle = \left\vert \phi _{4}\right\rangle \rangle $ hold when $\gamma _{x}=\omega $ $\left( \Omega=0\right) $, i.e., the system exhibits a second-order LEP.
\begin{figure}[hbt!]
	\centering
	\includegraphics[width=0.88\linewidth]{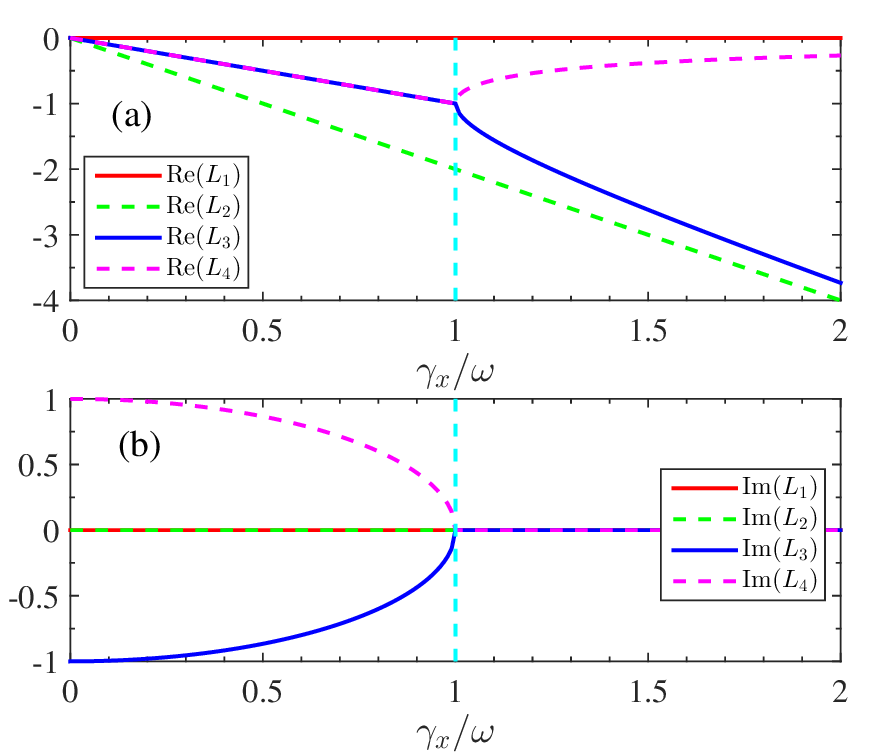}
	\caption{The eigenvalues of superpermatrix $\vec{L}(\omega )$  as a function of $\gamma _{x}/\omega $, in which (a) real part of eigenvalues; (b) imaginary part of eigenvalues. Here, we set $\omega=1 $ as a scale.}
	\label{fig1}
\end{figure}

In Fig.~\ref{fig1}, $L_i$ are plotted as a function of $\gamma _{x}/\omega $. 
One can see that $L_3$ (solid blue line) and $L_4$ (dotted pink line) are complex and have equal real parts and opposite imaginary parts when $\gamma_{x}/\omega<1$ (weak damping), while they become pure real number when $\gamma_{x}/\omega> 1$ (over damping), especially they are exactly equal at the LEP where $\gamma_{x}/\omega=1$ (critical damping).

If we define the dissipative spectral splitting $\Delta_{\omega}=L_{4}-L_{3}=2\Omega=2\sqrt{\gamma _{x}^{2}-\omega^{2}}$, the susceptibility of the splitting with respect to the parameter $\omega $ to be estimated, reads as 
\begin{equation}
\chi _{\omega }=\frac{\partial \Delta_{\omega} }{\partial \omega }=\left\vert \frac{%
	2\omega }{\sqrt{\gamma _{x}^{2}-\omega ^{2}}}\right\vert . \label{e55}
\end{equation}%
Obviously, $\chi _{\omega }$ diverges around the LEP, which seems to convey an intuition that if the frequency splitting of the Liouvillian spectrum can be accurately measured, one may get arbitrarily high measurement precision in the vicinity of the LEP. Later, we will employ DQFI as a means to illustrate that this pre-judgment is not necessarily accurate.

Based on Eqs.~(\ref{e41}) and (\ref{e54}), the dissipative generator $\widetilde{\Xi}_{\omega }$ is given
by 
\begin{equation}
\widetilde{\Xi}_{\omega }=\left[ 
\begin{array}{cccc}
0 & 0 & 0 & 0 \\ 
0 & \widetilde{\Xi}_{\omega }\left( 2,2\right)  & \widetilde{\Xi}_{\omega }\left(
2,3\right)  & 0 \\ 
0 & \widetilde{\Xi}_{\omega }\left( 3,2\right)  & \widetilde{\Xi}_{\omega }\left(
3,3\right)  & 0 \\ 
0 & 0 & 0 & 0%
\end{array}%
\right] , \label{e56}
\end{equation}%
in which
\begin{eqnarray*}
	\widetilde{\Xi}_{\omega }\left( 2,2\right)  &=&\frac{1}{\Omega ^{3}}\left[
	\frac{1}{2}\text{sinh}(2\Omega t)\gamma _{x}^{2}-t\omega ^{2}\Omega \right] ,
	\\
	\widetilde{\Xi}_{\omega }\left( 2,3\right)  &=&\frac{\gamma _{x}}{2\Omega ^{3}}%
	\left[ \Omega \left( 1-2it\omega \right) -\Omega \text{cosh}(2\Omega
	t)+i\omega \text{sinh}(2\Omega t)\right] , \\
	\widetilde{\Xi}_{\omega }\left( 3,2\right)  &=&\frac{\gamma _{x}}{2\Omega ^{3}}%
	\left[ -\Omega \left( 1+2it\omega \right) +\Omega \text{cosh}(2\Omega
	t)+i\omega \text{sinh}(2\Omega t)\right] , \\
	\widetilde{\Xi}_{\omega }\left( 3,3\right)  &=&\frac{1}{2}\left[ \frac{2t\omega ^{2}}{\Omega ^{2}}-\frac{\text{sinh%
		}(2\Omega t)\gamma _{x}^{2}}{\Omega ^{3}}\right].
\end{eqnarray*}%
In the current scenario, we have that
\begin{eqnarray}
\label{Eq65A}
\widetilde{\Theta}_{\omega } &=&\frac{\widetilde{\Xi}_{\omega }+\widetilde{\Xi}_{\omega
	}^{\dag }}{2}=\widetilde{\Xi}_{\omega }\left( 2,2\right) \left[ 
\begin{array}{cccc}
0 & 0 & 0 & 0 \\ 
0 & 1 & 0 & 0 \\ 
0 & 0 & -1 & 0 \\ 
0 & 0 & 0 & 0%
\end{array}%
\right] , \\
\widetilde{\Lambda}_{\omega } &=&\frac{i(\widetilde{\Xi}_{\omega }-\widetilde{\Xi}%
	_{\omega }^{\dag })}{2}=i\left[ 
\begin{array}{cccc}
0 & 0 & 0 & 0 \\ 
0 & 0 & \widetilde{\Xi}_{\omega }\left( 2,3\right) & 0 \\ 
0 & \widetilde{\Xi}_{\omega }\left( 3,2\right) & 0 & 0 \\ 
0 & 0 & 0 & 0%
\end{array}%
\right] .\label{Eq65B}
\end{eqnarray}%
The above equations holds for any $\Omega $, and  the forms of $\widetilde{\Xi}_{\omega }\left( 2,2\right)$, $\widetilde{\Xi}_{\omega }\left( 2,3\right)$ and $\widetilde{\Xi}_{\omega }\left(3,2\right)$ at the LEP ($\Omega=0$) are given by Eq.~(\ref{e57}). To attempt to clarify the
physical meanings of $\widetilde{\Theta}_{\omega }$ and $\widetilde{\Lambda}_{\omega
}$, we introduce the following orthogonal basis in the Liouville space
\begin{eqnarray}
|1\rangle \rangle &=&\left[ 
\begin{array}{c}
1 \\ 
0 \\ 
0 \\ 
0%
\end{array}%
\right] ,|2\rangle \rangle =\left[ 
\begin{array}{c}
0 \\ 
1 \\ 
0 \\ 
0%
\end{array}%
\right] ,  \notag \\
|3\rangle \rangle &=&\left[ 
\begin{array}{c}
0 \\ 
0 \\ 
1 \\ 
0%
\end{array}%
\right] ,|4\rangle \rangle =\left[ 
\begin{array}{c}
0 \\ 
0 \\ 
0 \\ 
1%
\end{array}%
\right] .
\end{eqnarray}%
Thus $\widetilde{\Theta}_{\omega }$ and $\widetilde{\Lambda}_{\omega }$ can be
rewritten as
\begin{eqnarray}
\widetilde{\Theta}_{\omega } &=&\widetilde{\Xi}_{\omega }\left( 2,2\right)
(|2\rangle \rangle \langle \langle 2|-|3\rangle \rangle \langle \langle 3|),
\\
\widetilde{\Lambda}_{\omega } &=&i\widetilde{\Xi}_{\omega }\left( 2,3\right)
|2\rangle \rangle \langle \langle 3|+i\widetilde{\Xi}_{\omega }\left( 3,2\right)
|3\rangle \rangle \langle \langle 2|,
\end{eqnarray}%
where $\widetilde{\Theta}_{\omega }$ and $\widetilde{\Lambda}_{\omega }$ are both
Hermitian, representing the population difference and energy level
transition between $|2\rangle \rangle $ and $|3\rangle \rangle $,
respectively. This indicates that an abstract non-unitary encoding process
is transformed into two physically explicit unitary encoding processes. Here, we need to point out that since $\widetilde{\Theta}_{\omega }$ and $\widetilde{\Lambda}_{\omega
}$ are Hermitian, introducing the orthogonal basis mentioned above to represent them is sufficient without the need for complicated biorthogonal bases.
 
According to the hyperbolic function in Eq.~(\ref{e56}),
whether $\Omega$ is a real or complex number has a significant impact on the temporal variation of $\widetilde{\Xi}_{\omega }$. In the case of $\omega >$ $\gamma _{x}$ ($\Omega $ is a complex number), $\widetilde{\Xi}_{\omega }$ tends to show the behavior of simple harmonic oscillations with time [corresponding to the second term in Eq.~(\ref{e41})] due to the dominant role of Hamiltonian in the dynamics. By comparison, in the case of $\omega <$ $\gamma _{x}
$ ($\Omega $ is a real number), $\widetilde{\Xi}_{\omega }$ towards the behavior of damped or gain oscillations with time owing to the dominant role of dissipation.  

At LEP, $\widetilde{\Xi}_{\omega }\Rightarrow \widetilde{\Xi}%
_{\omega }^{\text{LEP}}$ reads as 
\begin{equation}
\widetilde{\Xi}_{\omega }^{\text{LEP}}=\left[ 
\begin{array}{cccc}
0 & 0 & 0 & 0 \\ 
0 & \frac{2t^{3}\gamma _{x}^{2}}{3} & -\gamma _{x}t^{2}\left( 1-\frac{%
	i2\omega t}{3}\right) \  & 0 \\ 
0 & \gamma _{x}t^{2}\left( 1+\frac{i2\omega t}{3}\right)  & \frac{%
	-2t^{3}\gamma _{x}^{2}}{3} & 0 \\ 
0 & 0 & 0 & 0%
\end{array}%
\right] . \label{e57}
\end{equation}
The process of reducing Eq.~(\ref{e56}) to Eq.~(\ref{e57}) adopts the method of series expansion, namely when $x\rightarrow 0$, sinh$(x)\simeq x+x^{3}/6+o\left( x^{3}\right) $ and cosh$(x)\simeq
1+x^{2}/2+o\left( x^{3}\right) $ hold.  Of course, one can also use Eqs.~(\ref{e49}) and (\ref{e50}) to get the same result.

Assuming that the initial state of the system is the maximal superposition
state in Hilbert space, i.e., $\left\vert \Phi \left( 0\right) \right\rangle
=\left( \left\vert e\right\rangle +\left\vert g\right\rangle \right) /\sqrt{2%
}$, in which $\left\vert e\right\rangle $ and $\left\vert g\right\rangle $
are eigenstates of $\widehat{\sigma}_{z}$, satisfying $\widehat{\sigma}%
_{z}\left\vert e\right\rangle =\left\vert e\right\rangle $ and $\widehat{\sigma}%
_{z}\left\vert g\right\rangle =-\left\vert g\right\rangle $, then its expression in Liouville space reads as $\left\vert \widehat{\rho}_{s}\left(
0\right) \rangle \right\rangle =\frac{1}{2}\left[ 1,1,1,1\right] ^{\text{T}}$. The output state at time $t$ can be obtained by solving Eq.~(\ref{e52}), which takes the form,
\begin{equation}
\left\vert \widehat{\rho}_{s}\left( \omega ,t\right) \rangle \right\rangle _{N}=%
\frac{1}{\sqrt{\frac{1}{2}+2\left\vert \wp \right\vert ^{2}}}\left[ \frac{1}{%
	2},\wp ,\wp ^{\ast },\frac{1}{2}\right] ^{\text{T}}, \label{e58}
\end{equation}%
in which%
\[
\wp =\frac{e^{-\gamma _{x}t}}{2}\cosh (\Omega t)+\frac{\gamma _{x}e^{-\gamma
		_{x}t}}{2\Omega }\sinh (\Omega t)-\frac{i\omega e^{-\gamma _{x}t}}{2\Omega }%
\sinh (\Omega t).
\]%
For LEP, the state $|\widehat{\rho}_{s}^{\text{LEP}}\left( \omega ,t\right) \rangle \rangle _{N}$ can be obtained by replacing $\wp$ in Eq.~(\ref{e58}) with $\wp^{\text{LEP}}$ whose expression $\wp ^{\text{LEP}}=e^{-\gamma_{x}t}\left( 1+\gamma _{x}t-i\omega t\right) /2$ is derived with the assist of $\lim\limits_{\Omega\rightarrow 0}\sinh (\Omega t)/\Omega =t$ and $\lim\limits_{\Omega\rightarrow 0}\cosh (\Omega t)=1$.

Combining Eqs.~(\ref{e17}), (\ref{e56}) and (\ref{e58}), we obtain
\begin{eqnarray}
\widetilde{F}_{\omega }(t) &=&4\text{Cov}_{\left\vert \widehat{\rho}_{s}\left( \omega
	,t\right) \rangle \right\rangle _{N}}\left( \widetilde{\Xi}_{\omega }^{\dag },%
\widetilde{\Xi}_{\omega }\right)   \nonumber \\
&=&\frac{4}{\frac{1}{2}+2\left\vert \wp \right\vert ^{2}}\left[ 2\left\vert
\partial _{\omega }\wp \right\vert ^{2}-\frac{\left( \partial _{\omega
	}\left\vert \wp \right\vert ^{2}\right) ^{2}}{\frac{1}{2}+2\left\vert \wp
	\right\vert ^{2}}\right] ,\label{e59}
\end{eqnarray}%
where $\widetilde{F}_{\omega }(t)$ refers to the DQFI about $\omega $, which can quantify the estimation precision of $\omega$ in this open two-level system. 
Note that since Eq.~(\ref{e59}) contains the derivative operation of $\wp$ with respect to $\omega$, one cannot simply replace $\wp$ in the Eq.~(\ref{e59}) with $\wp ^{\text{LEP}}$ to obtain the DQFI at LEP, but use Eq.~(\ref{e17}) with $|\widehat{\rho}_{s}^{\text{LEP}}\left( \omega ,t\right) \rangle \rangle _{N}$ and $\widetilde{\Xi}_{\omega }^{\text{LEP}}$, that is,
\begin{eqnarray}
\widetilde{F}_{\omega }^{\text{LEP}}(t) &=&4\text{Cov}_{|\widehat{\rho}_{s}^{\text{LEP}%
	}\left( \omega ,t\right) \rangle \rangle _{N}}\left( \widetilde{\Xi}_{\omega }^{%
	\text{LEP}^{\dag }},\widetilde{\Xi}_{\omega }^{\text{LEP}}\right).\label{e60}
\end{eqnarray}
We do not give the specific form of $\widetilde{F}_{\omega }^{\text{LEP}}(t)$ here since it is tediously long. Nevertheless, it's still easy to find that $%
\widetilde{F}_{\omega }^{\text{LEP}}(t)$ does not diverge because $\wp ^{\text{LEP}}$ ($\widetilde{\Xi}_{\omega }^{\text{LEP}}$) is not a singular function (matrix) at the LEP, and Eq.~(\ref{e60}) is only a simple matrix multiplication operation.
This result indicates that the estimation of $\omega $ in the vicinity of LEP cannot obtain arbitrarily high precision. The physical reasons behind this are as follows. As mentioned earlier, in order to obtain any high sensitivity near LEP, it is necessary to accurately measure the frequency splitting of the Liuville spectrum. However, quantum state $|\widehat{\rho}_{s}^{\text{LEP}}\left( \omega ,t\right) \rangle \rangle _{N}$ at LEP does not explicitly contain $\Omega$,  which eventually leads to $\widetilde{F}_{\omega }^{\text{LEP}}(t)$ is a smooth function around the LEP. One can also think that this result originates from
the coalescence of the eigenvectors counteracts the susceptibility divergence at the LEP, and make the measurement precision becomes a smooth function of the parameters to be estimated \cite{chen2019sensitivity}. Physically, the correlation between different non-orthogonal eigenvectors can cause excess noise \cite{emile1998vectorial}, leading to the divergence in linewidth because the eigenvectors become maximally non-orthogonal (completely indistinguishable) at the LEP.  
Obviously, under the current model, the conclusion that  LEP cannot enhance the estimation precision is also held for CQFI $F_{\omega }^{\text{LEP}}(t)$. We will see this later.

As a remark, one reason exceptional-point-based sensors have been of recent interest is owing to the sensitivity of the spectral gap between the eigenvectors diverge at the exceptional point, which has been confirmed theoretically \cite{wiersig2014enhancing,wiersig2016sensors} and experimentally \cite{hodaei2017enhanced,chen2017exceptional}.  This has inspired researchers to explore whether the exceptional point can enhance measurement precision. Note also that pure sensitivity enhancement is not sufficient to enhance measurement precision \cite{wiersig2020prospects,chen2019sensitivity,langbein2018no,lau2018fundamental}. This is because the introduction of noise would also enhance the linewidth, and the final signal-to-noise ratio was not necessarily enhanced.  
Chen \emph{et al.} utilized CQFI to confirm that HEP does not offer a significant improvement in the estimation precision \cite{chen2019sensitivity}, which serves as a typical counterexample. Nevertheless, we also point out that in some cases, the enhanced excess noise can be surpassed by the improved response, ultimately resulting in enhanced measurement precision at the HEP \cite{kononchuk2022exceptional}. 
At present, we have obtained similar results to Chen \emph{et al.}'s work at the LEP. More importantly, our results are more convincing in explaining that exceptional point does not always improve measurement precision. 
Because the non-Hermitian Hamiltonian and its HEP involve only the coherent nonunitary evolution (i.e., energy loss or gain) without considering quantum jumps (i.e., decoherence), but the Liouvillian superoperator and its LEP include all dissipative effects. Hence the LEP more accurately describes the singularity of open systems compared to the HEP. 
In fact, the HEP can be interpreted as a semiclassical limit of the LEP \cite{minganti2019quantum}, but they two have fundamental differences at the quantum level. In addition, it is more natural and convenient to use DQFI to verify the singularity at LEP than CQFI in the Liouville space. However, it is still an open question under what physical mechanism the measurement precision near the exceptional point (HEP or LEP) can be improved, but this is beyond the scope of this paper.
\begin{figure}[hbt!]
	\centering
	\includegraphics[width=0.9\linewidth]{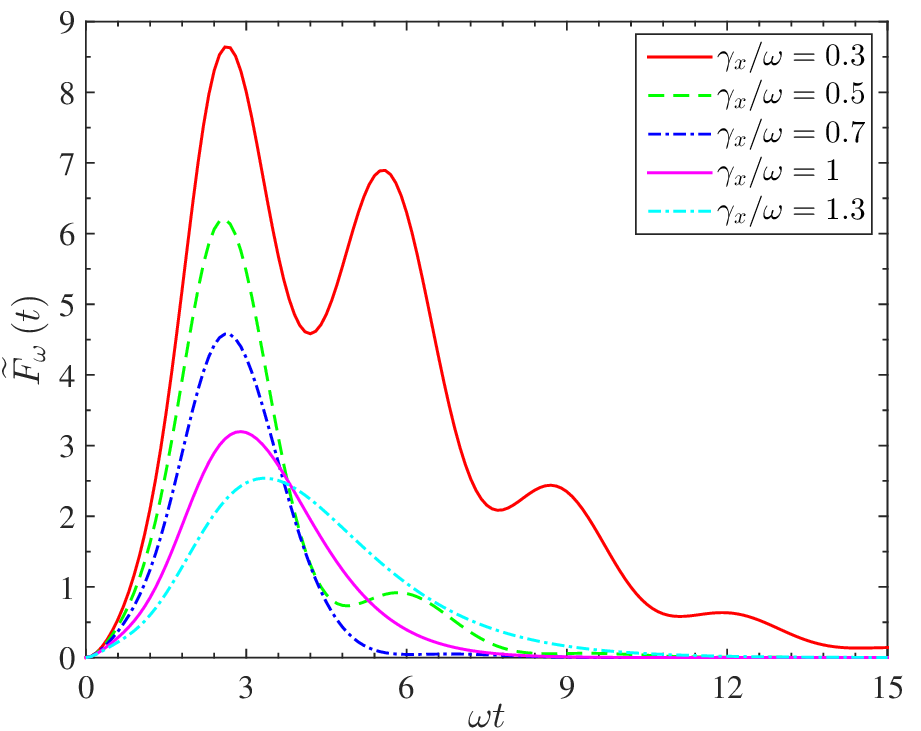}
	\caption{ The DQFI $\widetilde{F}_{\omega }\left( t\right)$ versus the evolution time for the different decay rates, where the pink solid line is drawn by Eq. (\ref{e60}), while others are drawn through Eq. (\ref{e59}). Here we take $\omega=1 $ as a scale.}
	\label{fig2}
\end{figure}

Fig.~\ref{fig2} shows the variation of  $\widetilde{F}_{\omega }\left( t\right)$ with the evolution time for different decay rates $\gamma_x$. One can see that, whatever $\gamma_x$ is, as time goes on $\widetilde{F}_{\omega }\left( t\right)$ gradually increases from its initial value of zero but finally decreases back to zero. In other words, the estimation error first decreases and finally tends to diverge with time.
This is because two physical processes are involved here: (i) internal Hamiltonian encoding quantum state; (ii) noise weakening the information of parameters to be estimated. The former makes $\widetilde{F}_{\omega }\left( t\right)$ increase, while the latter has the opposite effect. The competition between them determines the evolution of the estimation precision about $\omega$ with time. Under the long-term limit, the steady state of the system becomes a fully mixed state, which does not carry any information about $\omega$, i.e.,  $\widetilde{F}_{\omega }\left( t\right)=0$. In other words, the estimation error is infinite.

More specifically, one can see that the growth rate and the maximum value of $\widetilde{F}_{\omega }\left( t\right)$ depend strongly on the decay rate, that is, the smaller $\gamma_{x}$ leads to a larger growth rate and a larger maximum value owing to the dominance of Hamiltonian coding process over a longer time, and vice versa.
Particularly, we also proved numerically that  $\widetilde{F}_{\omega }\left( t\right)$ does not diverge at LEP (see the pink solid line). Moreover, one finds that with the increase of $\gamma_{x}$, the attenuation behavior of  $\widetilde{F}_{\omega }\left( t\right)$ changes from oscillatory decay to quasi-exponential attenuation. This may be explained in the following ways.
In the case of $\gamma_{x}\ll\omega$, $\Omega$ is a pure imaginary number with a large modulus. Based on the above discussion about Eqs. (\ref{e17}), (\ref{e41}) and (\ref{e42}), one can predict $\widetilde{F}_{\omega }\left( t\right)$ presents a grow with quasi-polynomial behavior of $t$ in a short time (see the first term in Eqs. (\ref{e41}) and (\ref{e42})), then tends to oscillatory behavior over time (see the second term in Eqs. (\ref{e41}) and (\ref{e42})), finally the oscillation attenuates to zero as the exponential decay factor of system state $|\widehat{\rho}_s(\omega, t)\rangle\rangle_N$ begins to dominate. 
When $\gamma_{x}$ is close to or greater than $\omega$, $\Omega$ is a pure imaginary number with a small modulus or a real number. Similarly, we can predict $\widetilde{F}_{\omega }\left( t\right)$ shows an increase with quasi-polynomial behavior of $t$ and then quickly towards a quasi-exponential decay behavior. Note also that the weak oscillations with time caused by a pure imaginary number with a small modulus may be covered by the exponential decay factor of the output state.  

Another interesting phenomenon is that in the range of $\gamma _{x}<\omega$, the increase of $\gamma _{x}$ leads to the reduction of the robustness of $\widetilde{F}_{\omega }\left( t\right)$ to noise, namely  $\widetilde{F}_{\omega }\left( t\right)$ decays to zero faster, while in the $\gamma _{x}\geq \omega $ region, improving $\gamma_{x}$ can strengthen the ability of  $\widetilde{F}_{\omega }\left( t\right)$ to resist noise (although the maximum  $\widetilde{F}_{\omega }\left( t\right)$ decreases), i.e., $\widetilde{F}_{\omega }\left( t\right)$ be able to last longer. The former is what we expect and is easy to understand. The latter is because under the strong-coupling regime between the system and the environment, the counter-rotating-wave  $\left[ \text{e.g., }\widehat{\sigma}_{+}\widehat{\rho}_{s}\left( \omega,t\right) 
\widehat{\sigma}_{+}\right] $ caused by flip-noise be able to enhance the robustness of $\widetilde{F}_{\omega }\left( t\right)$ resists the environment, thereby boosting the noisy quantum metrological performance \cite{zhang2022effects}.
\begin{figure}[hbt!]
	\centering
	\includegraphics[width=1.04\linewidth]{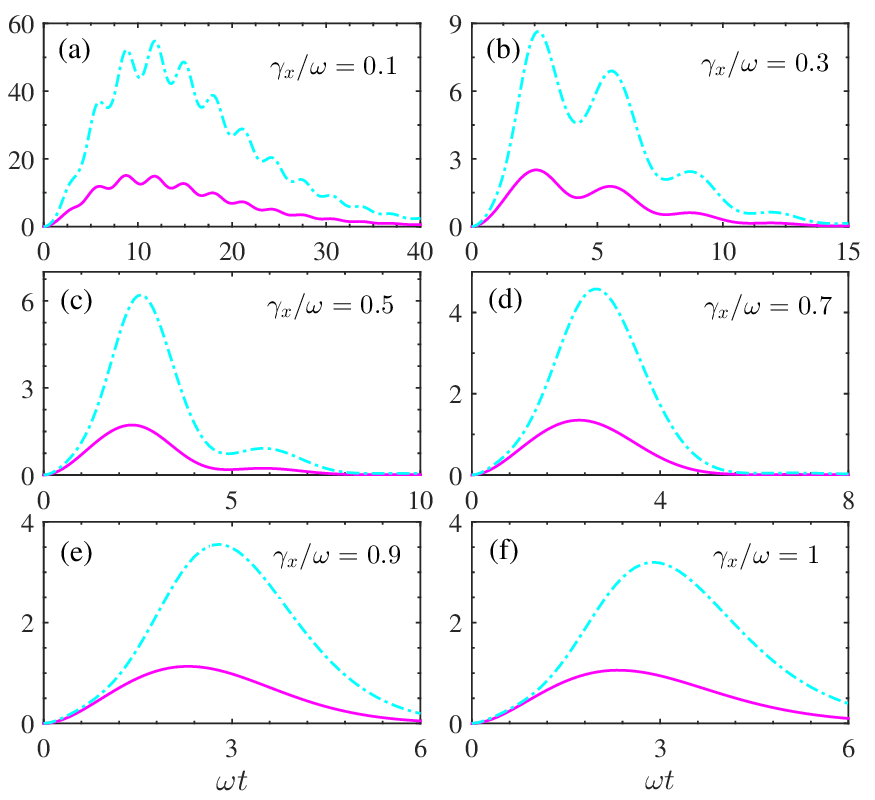}
	\caption{ The DQFI $\widetilde{F}_{\omega }\left( t\right)$ and CQFI  $F_{\omega }\left( t\right)$ as functions of the evolution time (scaled by $\omega^{-1}$) for different decay rates, in which the sky blue dotted line and the pink solid line represent $\widetilde{F}_{\omega }\left( t\right)$ and $F_{\omega }\left( t\right)$, respectively.}
	\label{fig3}
\end{figure}

One may be curious about the relationship between DQFI and CQFI  in the current model. In fact, in Ref. \cite{alipour2014quantum}, the inequality relationship between DQFI and CQFI has been given, but that is abstract, especially for mixed states. Here we will show them numerically, and then make some physically reasonable inferences.

First, one can get the form of the evolution state of the
system in Hilbert space by  $|\widehat{\rho}_{s}\left( \omega ,t\right) \rangle \rangle _{N}$ as follows \cite{casagrande2021analysis,minganti2019quantum,caves1999quantum}
\begin{equation}
\widehat{\rho}_{s}\left( \omega ,t\right) =\left[ 
\begin{array}{cc}
\frac{1}{2} & \wp \\ 
\wp ^{\ast } & \frac{1}{2}%
\end{array}%
\right]\,.
\end{equation}%
The basis-independent expression of CQFI for a single-qubit mixed state reads \cite{liu2020quantum}
\begin{eqnarray}
F_{\theta }\left[ \widehat{\rho}_{s}\left( t\right) \right]  &=&\text{Tr}\left[
\left( \partial _{\theta }\widehat{\rho}_{s}\left( t\right) \right) ^{2}\right] +%
\frac{1}{\det \left( \widehat{\rho}_{s}\left( t\right) \right) }  \nonumber \\
&&\text{Tr}\left[ \left( \widehat{\rho}_{s}\left( t\right) \partial _{\theta }%
\widehat{\rho}_{s}\left( t\right) \right) ^{2}\right],
\end{eqnarray}%
based on which, we obtain CQFI about $\omega $, the parameter to be estimated, i.e.,
\begin{equation}
F_{\omega }(t)=2\left\vert \partial _{\omega }\wp \right\vert ^{2}+\frac{%
	4\left( \wp \partial _{\omega }\wp ^{\ast }\right) ^{2}+4\left( \wp ^{\ast
	}\partial _{\omega }\wp \right) ^{2}+2\left\vert \partial _{\omega }\wp
	\right\vert ^{2}}{1-4\left\vert \wp \right\vert ^{2}}. \label{e63}
\end{equation}
Note that $F_{\omega }^{\text{LEP}}\left( t\right) $ at LEP is the limit result of taking $\Omega \rightarrow 0$ after the derivation of $\omega$ is completed by Eq.~(\ref{e63}). Below we show the numerical results directly.

Figure~\ref{fig3} presents the dynamics of the DQFI and CQFI versus evolution time for different decay rates, here the sky blue dotted and the pink solid lines denote $\widetilde{F}_{\omega }\left( t\right)$ and $F_{\omega }\left( t\right)$, respectively. One can see that $\widetilde{F}_{\omega }\left( t\right)$ is greater than $F_{\omega }\left( t\right)$ in the displayed parameter area (excluding the initial time and the final steady state).
This can be understood from the definition of QFI. CQFI is defined over all possible quantum measurements on the original system, while DQFI is obtained over all possible quantum measurements on the “extended system". The higher the dimensionality of the system, the more measurement options are available. More importantly, one finds that the overall line type profiles of  $\widetilde{F}_{\omega }\left( t\right)$ and $F_{\omega }\left( t\right)$ are very similar and have the same total number of extreme points. Note that the profile similarity here refers to the variation trend over time and the number of extreme points, not the values. This also indirectly confirms the comments in the last paragraph of Sec.~\ref{III}. 

In particular, when $\gamma_{x}$ is small, the difference in the location of their extreme points is very small (see Fig.~\ref{fig3}a). This is because for a pure state $\widetilde{F}_{\omega }\left( t\right) =2F_{\omega}\left( t\right) $ strictly holds. After the introduction of dissipation, the quantum state of the system will no longer be pure, which has a destructive effect on the similarity of the two line type profiles, and the stronger the noise the more significant the destruction (mainly reflected in the position difference of extreme points). The significance of this result is that when the noise is weak, we can replace the ideal metrology time of CQFI with that of DQFI. The latter may be much easier to solve than the former. This is because when the system dimension is large, the calculation of CSLD is very complex (requiring spectral decomposition of the density matrix), while the DSLD can be directly obtained through Eq.~(\ref{e15}). 
 
\section{Conclusion}\label{V}
In summary, beyond the framework for estimating the overall multiplicative factor of a Liouvillian superoperator, we have investigated the DQFI for general Liouvillian parameterized processes. By using the vectorization method, we circumvent the difficulty brought through the superoperator and derived the dissipative generator of the open system in the Liouville space. More importantly, we traced the origin of DQFI based on the dissipative generator.  The result showed that the dependence of eigenvalues and eigenvectors of Liouvillian supermatrix on estimation parameters is the key to the existence of DQFI. We considered estimating the transition frequency by measuring a two-level system with spin-flip noise as an example
to illustrate our theory. In particular, we employed DQFI to verify that the estimation precision near the LEP has not been significantly improved in this model. Further, we also found that DQFI and CQFI possess many similar features. Not limited to quantum metrology, DQFI may be similar to CQFI and will be connected with other aspects of the open quantum mechanics in the future, such as witness of entanglement and non-Markovianity, quantum thermodynamics, etc. Our study provides a new perspective on establishing connections between parameter estimation theory and other quantum effects for open systems from the vantage of vectorization. We also hope that the current work can provide certain theoretical guidance for those researchers who deeply study the properties and other potential applications of DQFI.

\section*{Acknowledgments}\label{VI}
This work was supported by the Innovation Program for
Quantum Science and Technology (No. 2021ZD0303200); the
National Key Research and Development Program of China
(No. 2016YFA0302001); the National Science Foundation of
China (Nos. 11974116, 12234014, and 11654005); the
Shanghai Municipal Science and Technology Major Project
(No. 2019SHZDZX01); the Fundamental Research Funds for
the Central Universities; the Chinese National Youth Talent
Support Program, and the Shanghai Talent program.
\appendix
\section{Properties and useful formulas of superoperator and vectorization} \label{AA}
Here, we lead readers to review the basic properties of superoperators and vectorization, which help to understand some of the derivations and calculations of
this paper. Below, we use the notation: the capital letters with a “$\wedge$” are operator, while the floral capital letters with a “$\wedge$” represent superoperator, as well as the double bracket  “$|$ $\rangle \rangle $” denotes a column vector in the Liouville space.

Superoperator: Mapping an operator to a linear mapping of a new operator, similar to the operator acting on a vector to generate a new vector. The two typical
superoperators in quantum mechanics are commutator $\mathcal{\widehat{A}}:=[ \widehat{A},\bullet]
$ and anti-commutator $\mathcal{\widehat{B}}:=\{ \widehat{B},\bullet\} $ (“$\bullet $” refers to the
operator being acted on), whose functions are $\mathcal{\widehat{A}}\widehat{G}=[ \widehat{A},%
\widehat{G}] =\widehat{A}\widehat{G}-\widehat{G}\widehat{A}$ and $\mathcal{\widehat{B}}\widehat{G}=\{ 
\widehat{B},\widehat{G}\} =\widehat{B}\widehat{G}+\widehat{G}\widehat{B}$, respectively. Here we adopt the traditional convention of always acting on the operator closest to the right of the “$\bullet $”. In addition, we also meet this superoperator in the master equation, namely $\mathcal{\widehat{C}}:=\widehat{A}\bullet \widehat{B}$ is such that $\mathcal{\widehat{C}}\widehat{G}=\widehat{A}\widehat{G}\widehat{B}.$ With these examples, we clearly see that the action of the superoperator is closed, i.e., the operator remains an operator after it is acted upon. Of course, there are also other types of super operators, which can be uniformly written as the right-hand side
action superoperator $R[\widehat{A}]\bullet :=\bullet \widehat{A}$ and the left-hand side
action superoperator $L[\widehat{A}]\bullet :=\widehat{A}\bullet $.

Vectorization: Mapping an operator (superoperator)  to a column vector (operator/matrix). This is similar to  “reshape” operations in Matlab and Python. The current work is to expand Hilbert space to Liouville space by vectorization method. For example, the operator $\widehat{A}$$\in$$\mathcal{H}$ in $N$-dimensional Hilbert space 
is mapped to the vector $|\widehat{A}\rangle \rangle =\left[ 
\begin{array}{cccc}
\widehat{A}\left( 1,:\right)  & \widehat{A} \left( 2,:\right)  & \cdots  & \widehat{A}%
\left( N,:\right) 
\end{array}%
\right] ^{\text{T}}$$\in$$\mathcal{H}^{\otimes 2}$ in the Liouville space \cite{minganti2019quantum,alipour2014quantum}, in which “T” in the upper right denotes the transpose and $A\left( N,:\right) $ represents the $N$th row of matrix $\widehat{A}$. This means that the
space dimension is enlarged by $N$ times. Note that, in some literatures \cite{liu2020quantum,vsafranek2018simple}, the vectorization of operator is defined as $|\widehat{A}%
\rangle \rangle =\left[ 
\begin{array}{cccc}
\widehat{A}\left( :,1\right) ^{\text{T}} & \widehat{A}\left( :,2\right) ^{\text{T}}
& \cdots  & \widehat{A}\left( :,N\right) ^{\text{T}}%
\end{array}%
\right] ^{\text{T}}$, where $\widehat{A}\left( :,N\right) $ refers to the $N$th
column of matrix $\widehat{A}$. Both definitions do not affect the final physical result. Here, we adopt the
first definition. A simple example, when $N$ $=2,$ we can get 
\begin{equation}
\widehat{A}=\left[ 
\begin{array}{cc}
a & b \\ 
c & d%
\end{array}%
\right] \underrightarrow{\text{vectorization}}\text{ }|\widehat{A}%
\rangle \rangle =\left[ 
\begin{array}{c}
a \\ 
b \\ 
c \\ 
d%
\end{array}%
\right].
\end{equation}
More generally, for an arbitrary linear operator $\widehat{\xi}$, we have that
\begin{equation}
\widehat{\xi}=\sum\limits_{i,j}\langle i|\widehat{\xi}|j\rangle |i\rangle \langle j|%
\text{ }\underrightarrow{\text{vectorization}} \text{ } |\widehat{\xi}\rangle
\rangle =\sum\limits_{i,j}\langle i|\widehat{\xi}|j\rangle |i\rangle |j\rangle ,
\end{equation}
where {$|i\rangle $} being the orthonormal basis in the Hilbert space $\mathcal{H}$.

The properties of vectorization have been well organized by S. Alipour \cite{alipour2014quantum},
i.e., 
\begin{subequations}
\begin{eqnarray}
\label{e80a}
\langle \langle \widehat{A}|\widehat{B}\rangle \rangle  &=&\text{Tr}[ \widehat{A}%
^{\dag }\widehat{B}] , \\
|\widehat{A}\widehat{B}\widehat{C}\rangle \rangle  &=&( \widehat{A}\otimes \widehat{C}^{%
	\text{T}}) |\widehat{B}\rangle \rangle , \\
|\widehat{A}\otimes \widehat{B}\otimes \widehat{C}\rangle \rangle  &=&|\widehat{A}\rangle
\rangle \otimes |\widehat{B}\rangle \rangle \otimes |\widehat{C}\rangle \rangle , \\
|[\widehat{A},\widehat{B}]\rangle \rangle  &=&( \widehat{A}\otimes \mathbb{1}-\mathbb{1}\otimes \widehat{A%
}^{\text{T}}) |\widehat{B}\rangle \rangle ,
\end{eqnarray}%
\end{subequations}
where $\langle \langle \widehat{A}|:=(|\widehat{A}\rangle \rangle )^{\dag }$; $\mathbb{1}$ represents the identity matrix with the same dimension as $\widehat{A}$. Please refer to Ref. \cite{alipour2014quantum} for detailed proof of the above formulas. By utilizing these properties, Eq.~(\ref{e9c}) is easily obtained.  In particular, based on Eq.~(\ref{e80a}) (Hilbert-Schmidt inner product), similar to the
Hermitian adjoint of operators, one can also define the Hermitian
adjoint of superoperators, namely \cite{minganti2019quantum,minganti2018out,carmichael2009statistical}
\begin{equation}
\label{e81}
\langle \langle \widehat{A}|\mathcal{\widehat{D}}\widehat{B}\rangle \rangle =\langle \langle \mathcal{\widehat{D}}^{\dag }\widehat{A}|\widehat{B}\rangle \rangle, 
\end{equation}%
where $\mathcal{\widehat{D}}$ is any superoperator. According to Eq.~(\ref{e81}), the following formulas can be verified
\begin{subequations}
\begin{eqnarray}
\langle \langle \widehat{C}|\mathcal{\widehat{A}}\widehat{B}\rangle \rangle  &=&\text{Tr}\left[ 
\widehat{C}^{\dag }\left[ \widehat{A},\widehat{B}\right] \right]   \notag \\
&=&\text{Tr}\left[ \left( \widehat{A}^{\dag }\widehat{C}-\widehat{C}\widehat{A}^{\dag
}\right) ^{\dag }\widehat{B}\right]   \notag \\
&=&\langle \langle \mathcal{\widehat{A}}^{\dag }\widehat{C}|\widehat{B}\rangle \rangle , \\
\langle \langle \widehat{C}|\mathcal{\widehat{B}}\widehat{A}\rangle \rangle  &=&\text{Tr}\left[ 
\widehat{C}^{\dag }\left\{ \widehat{B},\widehat{A}\right\} \right]   \notag \\
&=&\text{Tr}\left[ \left( \widehat{B}^{\dag }\widehat{C}+\widehat{B}^{\dag }\widehat{C}%
\right) ^{\dag }\widehat{A}\right]   \notag \\
&=&\langle \langle \mathcal{\widehat{B}}^{\dag }\widehat{C}|\widehat{A}\rangle \rangle , \\
\langle \langle \widehat{C}|\mathcal{\widehat{C}}\widehat{G}\rangle \rangle  &=&\text{Tr}\left[ 
\widehat{C}^{\dag }\widehat{A}\widehat{G}\widehat{B}\right]   \notag \\
&=&\text{Tr}\left[ \left( \widehat{A}^{\dag }\widehat{C}\widehat{B}^{\dag }\right)
^{\dag }\widehat{G}\right]   \notag \\
&=&\langle \langle \mathcal{\widehat{C}}^{\dag }\widehat{C}|\widehat{G}\rangle \rangle,
\end{eqnarray}%
\end{subequations}
where matrix trace in which the order of commutative matrix multiplication does not change has been used, i.e., Tr$[ \widehat{A}\widehat{B}\widehat{C}] =$ Tr$[ \widehat{C}\widehat{A}\widehat{B}%
] =$ Tr$[ \widehat{B}\widehat{C}\widehat{A}] $. One thus conclude that $\mathcal{\widehat{A}^{\dag }}:=[ \widehat{A}^{\dag },\bullet ] ,\mathcal{\widehat{B}^{\dag }}%
:=\{ \widehat{B}^{\dag },\bullet\} $ and $\mathcal{\widehat{C}}^{\dag }:=\widehat{A}^{\dag
}\bullet \widehat{B}^{\dag }.$  

On the other hand, operators and superoperators can be defined in Hilbert space or Liouville space. But we should notice that the operators and superoperators in Hilbert space correspond to the column vectors and operators in the Liouville space, respectively. This is the fundamental reason why this paper introduces vectorization technology to simplify the processing of dissipative systems.

\section{Basic properties of non-Hermitian matrix and its adjoint} \label{AB}
Considering some readers are not familiar with the relevant contents of
non-Hermitian quantum mechanics. For pedagogical reasons, let's also review
the main aspects of non-Hermitian matrix and its adjoint. In order to be more in line with the theme of this paper, the following discussion about this part is conducted in Liouville space. Suppose that a non-Hermitian matrix in the Liouville space reads
\begin{equation}
\label{e83}
\vec{L}_{\text{eff}}=\vec{K}-i\vec{\Gamma},
\end{equation}%
where $\vec{K}$  and $\vec{\Gamma}$ are Hermitian matrices that satisfy $%
\vec{K}^{\dag }=\vec{K}$ and $\vec{\Gamma}^{\dag }=\vec{\Gamma}$,
respectively. What one needs to point out here is that the use of “$\rightarrow $” is to be consistent with the main text. In fact, it represents the operator in Liouville space. Obviously, $\vec{L}_{\text{eff}}^{\dag }\neq \vec{L}_{\text{eff%
}}$, hence $\vec{L}_{\text{eff}}$ is a non-Hermitian. Note that
Eq.~(\ref{e83}) is general, as any non-Hermitian matrix can always be split
into a combination of two Hermitian matrices. For instance, for the
non-Hermitian matrix $\vec{L}_{\text{eff}}=\vec{Q}$, the corresponding
ones are $\vec{K}=( \vec{Q}+\vec{Q}^{\dag }) /2$ and $\vec{\Gamma}=i(%
\vec{Q}-\vec{Q}^{\dag })/2$. 

Let $\left\{ \left\vert E_{n}\right\rangle \rangle\right\} $ and $\left\{
E_{n}\right\} $ be the right eigenvectors and eigenvalues of $\vec{L}_{\text{%
		eff}}$, respectively, satisfying eigenequation
\begin{subequations}	
	\label{e84}
\begin{eqnarray}
\vec{L}_{\text{eff}}\left\vert E_{n}\right\rangle\rangle  &=&E_{n}\left\vert
E_{n}\right\rangle\rangle , \\
\left\langle\langle E_{n}\right\vert \vec{L}_{\text{eff}}^{\dag } &=&\left\langle\langle
E_{n}\right\vert E_{n}^{\ast },
\end{eqnarray}%
\end{subequations}
where $E_{n}$ not necessarily be a real number owing to $\vec{L}_{\text{eff}}
$ is non-Hermitian. Unless stated otherwise,  we first assume that $\left\{
E_{n}\right\} $ does not degenerate. Based on Eqs.~(\ref{e84}), for $m\neq n$ one can easily obtain 
\begin{subequations}
\label{e85}
\begin{eqnarray}
\left\langle\langle E_{m}\right\vert \vec{L}_{\text{eff}}\left\vert
E_{n}\right\rangle\rangle  &=&E_{n}\left\langle\langle E_{m}|E_{n}\right\rangle\rangle , \\
\left\langle\langle E_{m}\right\vert \vec{L}_{\text{eff}}^{\dag }\left\vert
E_{n}\right\rangle\rangle  &=&E_{m}^{\ast }\left\langle\langle E_{m}|E_{n}\right\rangle\rangle .
\end{eqnarray}%
\end{subequations}
Further, we can get by Eqs.~(\ref{e85})
\begin{eqnarray}
\label{e86}
\left\langle \langle E_{m}|E_{n}\right\rangle\rangle  &=&2\frac{\left\langle\langle
	E_{m}\right\vert \vec{K}\left\vert E_{n}\right\rangle\rangle }{E_{n}+E_{m}^{\ast }}
\notag \\
&=&2i\frac{\left\langle\langle E_{m}\right\vert \vec{\Gamma}\left\vert
	E_{n}\right\rangle\rangle }{E_{m}^{\ast }-E_{n}}.
\end{eqnarray}
Therefore,$\ \left\vert E_{n}\right\rangle\rangle $ and $\left\vert
E_{m}\right\rangle\rangle $ may not necessarily be orthogonal to each other. 

Particularly, the non-orthogonality of eigenvectors leads to imperfect
projection techniques \cite{brody2013biorthogonal}. To this end, one can introduce the eigenvectors $%
\left\{ \left\vert H_{n}\right\rangle\rangle \right\} $ of Hermitian adjoint $\vec{L%
}_{\text{eff}}^{\dag }$ (them can also be called the left eigenvectors of $%
\vec{L}_{\text{eff}}$), i.e., 
\begin{subequations}
\label{e87}	
\begin{eqnarray}
\vec{L}_{\text{eff}}^{\dag }\left\vert H_{n}\right\rangle \rangle
&=&H_{n}\left\vert H_{n}\right\rangle\rangle , \\
\left\langle\langle H_{n}\right\vert \vec{L}_{\text{eff}} &=&\left\langle\langle
H_{n}\right\vert H_{n}^{\ast }.
\end{eqnarray}%
\end{subequations}
Similarly, one can also obtain
\begin{eqnarray}
\label{e88}
\left\langle\langle H_{m}|H_{n}\right\rangle\rangle  &=&2\frac{\left\langle\langle
	H_{m}\right\vert \vec{K}\left\vert H_{n}\right\rangle\rangle }{H_{n}+H_{m}^{\ast }}
\notag \\
&=&2i\frac{\left\langle\langle H_{m}\right\vert \vec{\Gamma}\left\vert
	H_{n}\right\rangle\rangle }{H_{n}-H_{m}^{\ast }}.
\end{eqnarray}%
According to Eqs.~(\ref{e84}) and~(\ref{e87}), the following result holds
\begin{subequations}
	\label{e89}
\begin{eqnarray}
\label{e89a}
\left\langle\langle H_{m}\right\vert \vec{L}_{\text{eff}}\left\vert
E_{n}\right\rangle\rangle   &=&E_{n}\left\langle\langle H_{m}|E_{n}\right\rangle\rangle , \\
\left\langle\langle H_{m}\right\vert \vec{L}_{\text{eff}}\left\vert
E_{n}\right\rangle\rangle  &=&H_{m}^{\ast }\left\langle\langle H_{m}|E_{n}\right\rangle\rangle.\label{e89b}
\end{eqnarray}
\end{subequations}
Eq.~(\ref{e89a}) minus Eq.~(\ref{e89b}) to obtain $\left\langle\langle H_{m}|E_{n}\right\rangle\rangle =\delta
_{mn}\left\langle\langle H_{m}|E_{n}\right\rangle\rangle $ owing to $E_{n}\neq H_{m}^{\ast
}$ $\left( n\neq m\right) $ and $E_{n}=H_{n}^{\ast }$ (note that $E_{n}$
must be equal to an element in $\left\{ H_{m}^{\ast }\right\} $. Without loss of generality, here we have labeled the states
such that $E_{n}=H_{n}^{\ast }$), which means space $\left\{ \left\vert
E_{n}\right\rangle\rangle \right\} $ and its dual space $\left\{ \left\vert
H_{n}\right\rangle\rangle \right\} $ show biorthogonality \cite{brody2013biorthogonal,curtright2007biorthogonal,chang2013biorthogonal}.

On the other hand, although $\left\{ \left\vert E_{n}\right\rangle\rangle \right\} $
are not orthogonal, they are independent of each other. It is easy to prove
that $C_{n}=0$ $\left( \text{for}~\forall~n\right) $ in $\sum\nolimits_{n}C_{n}\left\vert E_{n}\right\rangle =0$ through $\left\langle\langle H_{m}|E_{n}\right\rangle\rangle
=\delta _{mn}\left\langle\langle H_{m}|E_{n}\right\rangle\rangle $, that is $\left\{
\left\vert E_{n}\right\rangle\rangle \right\} $ are linearly independent \cite{brody2013biorthogonal,curtright2007biorthogonal,chang2013biorthogonal}. As a
result, $\left\{ \left\vert E_{n}\right\rangle\rangle \right\} $  can forms a
complete non-orthogonal basis. Therefore, an arbitrary matrix $\vec{O}$
can be represented as 
\begin{equation}
	\label{e90}
\vec{O}=\sum\limits_{n,m}O_{mn}\left\vert E_{m}\right\rangle\rangle \left\langle\langle 
E_{n}\right\vert ,
\end{equation}%
where $O_{mn}=\left\langle\langle E_{m}\right\vert \vec{O}\left\vert
E_{n}\right\rangle\rangle $ and $\left\{ \left\vert E_{n}\right\rangle\rangle \right\} $
has been normalized (a familiar example is the coherent state
representation).  The similar results hold true for $\left\{ \left\vert
H_{n}\right\rangle\rangle \right\} $. 

More interesting, one can find that the following formula holds
\begin{equation}
	\label{e91}
\sum\limits_{n}\vec{T}_{n}=\sum\limits_{n}\vec{J}_{n}=\mathbb{1}_{n},
\end{equation}%
with
\begin{equation}
	\label{e92}
\vec{T}_{n}=\frac{\left\vert E_{n}\right\rangle\rangle \left\langle\langle
	H_{n}\right\vert }{\left\langle\langle H_{n}|E_{n}\right\rangle\rangle },\vec{J}_{n}=\frac{%
	\left\vert H_{n}\right\rangle\rangle \left\langle\langle E_{n}\right\vert }{\left\langle\langle
	E_{n}|H_{n}\right\rangle\rangle },
\end{equation}%
where $\{ \vec{T}_{n}\} $ and $\{ \vec{J}_{n}\} $ can
be regarded as complex projection operators in Liouville space. Eq.~(\ref{e91}) is similar in form to a
sup-completeness relationship, which can be considered as a biorthogonal
basis composed of $\left\{ \left\vert E_{n}\right\rangle\rangle \right\} $ and $%
\left\{ \left\vert H_{n}\right\rangle\rangle \right\} $ in non-Hermitian physics.
In practice, to ensure that the probabilistic interpretation of measurement
output under biorthogonal basis is consistent with traditional quantum
mechanics, one can use the following convention \cite{brody2013biorthogonal,curtright2007biorthogonal}
\begin{equation}
	\label{e93}
\left\langle\langle H_{n}|E_{n}\right\rangle\rangle =\left\langle\langle E_{n}|H_{n}\right\rangle\rangle
=1.
\end{equation}%
This convention has been widely used in the field of quantum chemistry and
it brings great convenience. In this case, Eq.~(\ref{e92}) is rewritten as
\begin{equation}
	\label{e94}
\vec{T}_{n}=\left\vert E_{n}\right\rangle\rangle \left\langle\langle H_{n}\right\vert ,%
\vec{J}_{n}=\left\vert H_{n}\right\rangle\rangle \left\langle\langle E_{n}\right\vert.
\end{equation}
Moreover, for simplicity, one can assume that $%
\left\langle\langle E_{n}|E_{n}\right\rangle\rangle =1$ for all $n$, the corresponding $%
\left\langle\langle H_{n}|H_{n}\right\rangle\rangle $, are determined by Eq.~(\ref{e93}). Under
biorthogonal basis $\left\{ \left\vert E_{n}\right\rangle\rangle ,\left\vert
H_{n}\right\rangle\rangle \right\} $, a arbitrary matrix $\vec{O}$  can be
expressed  as
\begin{eqnarray}
	\label{e95}
\vec{O} &=&\sum\limits_{n,m}O_{1,nm}\left\vert E_{n}\right\rangle\rangle
\left\langle\langle H_{m}\right\vert   \notag \\
&=&\sum\limits_{n,m}O_{2,nm}\left\vert H_{n}\right\rangle\rangle \left\langle\langle 
E_{m}\right\vert, 
\end{eqnarray}%
where matrix elements $O_{1,nm}$ $=\left\langle\langle H_{n}\right\vert \vec{O}\left\vert
E_{m}\right\rangle\rangle $ and $O_{2,nm}$ $=\left\langle\langle E_{n}\right\vert \vec{O}%
\left\vert H_{m}\right\rangle\rangle$, they are the results of using projection operators $\vec{T}_{n}$ and $\vec{J}_{n}$, respectively.

In the case of  $\left\{E_{n}\right\} $ degeneracy, some of the above results need special treatment. A typical situation is the occurrence of exceptional points, where at least two the eigenvalues and corresponding eigenvectors be merged simultaneously \cite{wiersig2014enhancing,wiersig2016sensors,wiersig2020prospects,chen2019sensitivity,langbein2018no,lau2018fundamental}. In this case, Eqs.~(\Ref{e86}) and~(\Ref{e88}) may have a denominator of $0$ under certain $n$, leading to the calculation
of $\left\langle\langle E_{m}|E_{n}\right\rangle\rangle$ and $\left\langle\langle H_{m}|H_{n}\right\rangle\rangle$ needs to be careful. Moreover,  $\left\{ \left\vert E_{n}\right\rangle \rangle\right\} $  and  $\left\{ \left\vert H_{n}\right\rangle \rangle\right\} $ are no longer complete due to the combination of eigenvectors. At this time, the construction of biorthogonal basis needs to use the method provided by  Ref.~\cite{sarandy2005adiabatic}.

\section{Spectral decomposition formulas of DQFI and DSLD} \label{AC}
In order to complete this paper, here we present the formula for calculating the spectral decomposition of DQFI and DSLD base on the density matrix.  Assuming that the spectral
decompostion of density matrix $\widehat{\rho}_{\theta }$ takes the form 
\begin{equation}
\label{e96}
\widehat{\rho}_{\theta }=\sum_{k=1}^{m}p_{k}\left\vert \psi _{k}\right\rangle
\left\langle \psi _{k}\right\vert ,
\end{equation}%
here $p_{k}>0$ $\left( \left\vert \psi _{k}\right\rangle \right) $ is $k$th
eigenvalue (eigenstate) of $\widehat{\rho}_{\theta}$; $m$ denotes the support dimension
of $\widehat{\rho}_{\theta}$. As $\widehat{\rho}_{\theta}$ is a Hermitian operator, hence $\left\{ \left\vert \psi _{k}\right\rangle \right\} $ satisfying completeness
\begin{equation}
\label{e97}
\sum_{k=1}^{d}\left\vert \psi _{k}\right\rangle \left\langle \psi
_{k}\right\vert =\mathbb{1}_{d},
\end{equation}%
where $d=$ dim$\left( \widehat{\rho}_{\theta }\right) $ is dimension
of $\widehat{\rho}_{\theta}$ and $m\leq d$ (the equal sign indicates that $\widehat{\rho}_{\theta}$ is full rank matrix). Known
CSLD $\widehat{\mathfrak{M}}_{\theta }$ about estimated parameter $\theta $, defined in a way that
\begin{equation}
\label{e98}
\frac{\partial \widehat{\rho}_{\theta }}{\partial \theta }=\frac{1}{2}\left( 
\widehat{\rho}_{\theta }\widehat{\mathfrak{M}}_{\theta }+\widehat{\mathfrak{M}}_{\theta }\widehat{\rho}_{\theta
}\right) .
\end{equation}%
Under $\left\{ \left\vert \psi _{k}\right\rangle \right\} $ basis, one can
obtain
\begin{equation}
\label{e99}
( \widehat{\mathfrak{M}}_{\theta }) _{kj}=\frac{\partial _{\theta }p_{k}}{p_{k}}%
\delta _{kj}-\frac{2\left( p_{k}-p_{j}\right) }{p_{k}+p_{j}}\left\langle
\psi _{k}\right\vert \partial _{\theta }\psi _{j}\rangle ,
\end{equation}%
where matrix element $(\widehat{\mathfrak{M}}_{\theta }) _{kj}=\left\langle \psi
_{k}\right\vert \widehat{\mathfrak{M}}_{\theta }\left\vert \psi _{j}\right\rangle $; $p_{k}\neq 0$ and $p_{k}+p_{j}\neq 0$.  Note
that the form of $\widehat{\mathfrak{M}}_{\theta }$ is not unique for non-full rank density
matrix $\widehat{\rho}_{\theta }$, but this not affect the value of CQFI \cite{liu2020quantum}. This is because when calculating CQFI, there is no case in which the indexes $k$ and $j$ are greater than $m$ at the same time in $( \widehat{\mathfrak{M}}_{\theta }) _{kj}$. For convenience, it can be assumed that $( \widehat{\mathfrak{M}}_{\theta }) _{kj}=0$ for $i,j>m$. Substituting Eqs.~(\ref{e96})-(\ref{e99})  into Eqs. (\ref{e2c}), the spectral decomposition formula for CQFI is obtained, i.e., 
\begin{widetext}
\begin{equation}
\label{e100}
F_{\theta }=\sum_{k=1}^{m}\frac{\left[ \partial _{\theta }p_{k}\right] ^{2}}{%
	p_{k}}+4\sum_{k=1}^{m}p_{k}\left\langle \partial _{\theta }\psi
_{k}\right\vert \partial _{\theta }\psi _{k}\rangle
-\sum_{k=1}^{m}\sum_{j=1}^{m}\frac{8p_{k}p_{j}}{p_{k}+p_{j}}\left\vert
\left\langle \psi _{k}\right\vert \partial _{\theta }\psi _{j}\rangle
\right\vert ^{2}.
\end{equation}
This expression is applicable to the calculation of QFI of density matrix with arbitrary rank.

According to Ref.~\cite{alipour2014quantum}, one know that
\begin{equation}
\label{e101}
\widetilde{F}_{\theta }=\frac{2}{\text{Tr}\left[ \widehat{\rho}_{\theta }^{2}\right] 
}\left[ \text{Tr}\left[ \widehat{\rho}_{\theta }\widehat{\mathfrak{M}}_{\theta }\widehat{\rho}%
_{\theta }\widehat{\mathfrak{M}}_{\theta }\right] +\text{Tr}\left[ \widehat{\rho}_{\theta }^{2}%
\widehat{\mathfrak{M}}_{\theta }^{2}\right] -2\frac{\left( \text{Tr}\left[ \widehat{\rho}%
	_{\theta }^{2}\widehat{\mathfrak{M}}_{\theta }\right] \right) ^{2}}{\text{Tr}\left[ \widehat{%
		\rho}_{\theta }^{2}\right] }\right] .
\end{equation}%
Obviously, in order to calculate the spectral decomposition form of DQFI $\widetilde{F}_{\theta }$, one needs to further evaluate Tr$\left[ 
\widehat{\rho}_{\theta }^{2}\right]$, Tr$[ \widehat{\rho}_{\theta }\widehat{\mathfrak{M}}_{\theta }\widehat{\rho}_{\theta }\widehat{\mathfrak{M}}_{\theta }]$, Tr$[ \widehat{\rho}%
_{\theta }^{2}\widehat{\mathfrak{M}}_{\theta }^{2}] $ and Tr$[ \widehat{\rho}%
_{\theta }^{2}\widehat{\mathfrak{M}}_{\theta }] $, i.e., 
\begin{subequations}
	\label{e102}
\begin{eqnarray}
\text{Tr}\left[ \widehat{\rho}_{\theta }^{2}\right]  &=&\sum_{k=1}^{m}p_{k}^{2},%
\text{Tr}\left[ \widehat{\rho}_{\theta }^{2}\widehat{\mathfrak{M}}_{\theta }\right]
=\sum_{k=1}^{m}p_{k}\left[ \partial _{\theta }p_{k}\right],  \\
\text{Tr}\left[ \widehat{\rho}_{\theta }\widehat{\mathfrak{M}}_{\theta }\widehat{\rho}_{\theta }%
\widehat{\mathfrak{M}}_{\theta }\right]  &=&\sum_{k=1}^{m}\left[ \partial _{\theta }p_{k}%
\right] ^{2}+\sum_{k=1}^{m}\sum_{j=1}^{m}p_{k}p_{j}\frac{4\left(
	p_{k}-p_{j}\right) ^{2}}{\left( p_{k}+p_{j}\right) ^{2}}\left\vert
\left\langle \psi _{k}\right\vert \partial _{\theta }\psi _{j}\rangle
\right\vert ^{2}, \\
\text{Tr}\left[ \widehat{\rho}_{\theta }^{2}\widehat{\mathfrak{M}}_{\theta }^{2}\right] 
&=&\sum_{k=1}^{m}\left[ \partial _{\theta }p_{k}\right] ^{2}+\sum_{k=1}^{m}%
\sum_{j=1}^{d}\frac{4p_{k}^{2}\left( p_{k}-p_{j}\right) ^{2}}{\left(
	p_{k}+p_{j}\right) ^{2}}\left\vert \left\langle \psi _{k}\right\vert
\partial _{\theta }\psi _{j}\rangle \right\vert ^{2}.
\end{eqnarray}%
\end{subequations}
In the derivation of the above formulas, we have used Eqs.~(\ref{e96}),~(\ref{e97}) and~(\ref{e99}). Moreover, $( \widehat{\mathfrak{M}}_{\theta }) _{kj}$ with both $k$ and $j$ greater than $m$ does not contribute to the calculation of the  Eqs.~(\ref{e102}).  Therefore, using Eq.~(\Ref{e99}) to calculate $\widetilde{F}_{\theta }$ is convincing. In addition,
\begin{eqnarray}
\label{e103}
&&\sum_{k=1}^{m}\sum_{j=1}^{d}\frac{4p_{k}^{2}\left( p_{k}-p_{j}\right) ^{2}%
}{\left( p_{k}+p_{j}\right) ^{2}}\left\vert \left\langle \psi
_{k}\right\vert \partial _{\theta }\psi _{j}\rangle \right\vert ^{2}  \notag
\\
&=&\sum_{k=1}^{m}\sum_{j=1}^{m}\frac{4p_{k}^{2}\left( p_{k}-p_{j}\right) ^{2}%
}{\left( p_{k}+p_{j}\right) ^{2}}\left\vert \left\langle \psi
_{k}\right\vert \partial _{\theta }\psi _{j}\rangle \right\vert
^{2}+\sum_{k=1}^{m}\sum_{j=m+1}^{d}4p_{k}^{2}\left\vert \left\langle \psi
_{k}\right\vert \partial _{\theta }\psi _{j}\rangle \right\vert ^{2}  \notag
\\
&=&\sum_{k=1}^{m}\sum_{j=1}^{m}\frac{4p_{k}^{2}\left( p_{k}-p_{j}\right) ^{2}%
}{\left( p_{k}+p_{j}\right) ^{2}}\left\vert \left\langle \psi
_{k}\right\vert \partial _{\theta }\psi _{j}\rangle \right\vert
^{2}+\sum_{k=1}^{m}4p_{k}^{2}\left\langle \partial _{\theta }\psi
_{k}\right\vert \partial _{\theta }\psi _{k}\rangle
-\sum_{k=1}^{m}\sum_{j=1}^{m}4p_{k}^{2}\left\vert \left\langle \psi
_{k}\right\vert \partial _{\theta }\psi _{j}\rangle \right\vert ^{2}  \notag
\\
&=&\sum_{k=1}^{m}4p_{k}^{2}\left\langle \partial _{\theta }\psi
_{k}\right\vert \partial _{\theta }\psi _{k}\rangle
-\sum_{k=1}^{m}\sum_{j=1}^{m}\frac{8p_{k}p_{j}\left(
	p_{k}^{2}+p_{j}^{2}\right) }{\left( p_{k}+p_{j}\right) ^{2}}\left\vert
\left\langle \psi _{k}\right\vert \partial _{\theta }\psi _{j}\rangle
\right\vert ^{2},
\end{eqnarray}%
where we have used 
\begin{subequations}
\begin{eqnarray}
\sum_{j=m+1}^{d}\left\vert \psi _{j}\right\rangle \left\langle \psi
_{j}\right\vert  &=&\mathbb{1}_{d}-\sum_{j=1}^{m}\left\vert \psi _{j}\right\rangle
\left\langle \psi _{j}\right\vert ,\left\langle \psi _{k}\right\vert
\partial _{\theta }\psi _{j}\rangle =-\left\langle \partial _{\theta }\psi
_{k}\right\vert \psi _{j}\rangle , \\
\sum_{k=1}^{m}\sum_{j=1}^{m}4p_{k}^{2}\left[ \frac{\left( p_{k}-p_{j}\right)
	^{2}}{\left( p_{k}+p_{j}\right) ^{2}}-1\right] \left\vert \left\langle \psi
_{k}\right\vert \partial _{\theta }\psi _{j}\rangle \right\vert ^{2}
&=&\sum_{k=1}^{m}\sum_{j=1}^{m}4p_{j}^{2}\left[ \frac{\left(
	p_{k}-p_{j}\right) ^{2}}{\left( p_{k}+p_{j}\right) ^{2}}-1\right] \left\vert
\left\langle \psi _{k}\right\vert \partial _{\theta }\psi _{j}\rangle
\right\vert ^{2}.
\end{eqnarray}
\end{subequations}
Substituting Eqs.~(\Ref{e102})-(\Ref{e103}) into Eq.~(\Ref{e101}), we finally obtain
\begin{equation}
\label{e105}
\widetilde{F}_{\theta }=4\sum_{k=1}^{m}\frac{\left[ \partial _{\theta }p_{k}%
	\right] ^{2}}{\sum_{k=1}^{m}p_{k}^{2}}+\sum_{k=1}^{m}\frac{8p_{k}^{2}}{%
	\sum_{k=1}^{m}p_{k}^{2}}\left\langle \partial _{\theta }\psi _{k}\right\vert
\partial _{\theta }\psi _{k}\rangle -\sum_{k=1}^{m}\sum_{j=1}^{m}\frac{%
	8p_{k}p_{j}}{\sum_{k=1}^{m}p_{k}^{2}}\left\vert \left\langle \psi
_{k}\right\vert \partial _{\theta }\psi _{j}\rangle \right\vert ^{2}-4\frac{%
	\sum_{k=1}^{m}\sum_{j=1}^{m}p_{k}p_{j}\left[ \partial _{\theta }p_{k}\right] %
	\left[ \partial _{\theta }p_{j}\right] }{\sum_{k=1}^{m}p_{k}^{2}%
	\sum_{j=1}^{m}p_{j}^{2}}.
\end{equation}%
For the pure state, namely $\widehat{\rho}_{\theta }\Rightarrow \left\vert \psi \right\rangle \left\langle
\psi \right\vert $,  it's easy to get
\begin{equation}
\label{e106}
F_{\theta }=4\left( \left\langle \partial _{\theta }\psi \right\vert
\partial _{\theta }\psi \rangle -\left\vert \left\langle \psi \right\vert
\partial _{\theta }\psi \rangle \right\vert^{2} \right),\widetilde{F}_{\theta
}=8\left( \left\langle \partial _{\theta }\psi \right\vert \partial _{\theta
}\psi \rangle -\left\vert \left\langle \psi \right\vert \partial _{\theta
}\psi \rangle \right\vert^{2} \right).
\end{equation}
Hence $\widetilde{F}_{\theta }=2F_{\theta }$ holds for the pure state, which originates from the increase of the spatial dimension.  

Next, we derive the spectral decomposition form of DSLD $\widetilde{\mathfrak{M}}_{\theta } $. Known formula \cite{alipour2014quantum}
\begin{equation}
\label{e107}
\widetilde{\mathfrak{M}}_{\theta } =\widehat{\mathfrak{M}}_{\theta }\otimes \mathbb{1}_{d}+\mathbb{1}_{d}\otimes \widehat{\mathfrak{M}}_{\theta
}^{\text{T}}-2\frac{\text{Tr}\left[ \widehat{\rho}_{\theta }^{2}\widehat{\mathfrak{M}}_{\theta }%
	\right] }{\text{Tr}\left[ \widehat{\rho}_{\theta }^{2}\right] }.
\end{equation}%
Similarly, by utilizing Eqs.~(\ref{e102}), one can obtain
\begin{equation}
\label{e107}
\widetilde{\mathfrak{M}}_{\theta } 
=\sum_{k=1}^{d}\sum_{j=1}^{d}( \widehat{\mathfrak{M}}_{\theta }) _{kj} \left\vert \psi _{k}\right\rangle \left\langle \psi _{j}\right\vert\otimes
\mathbb{1}_{d}+\mathbb{1}_{d}\otimes \sum_{k=1}^{d}\sum_{j=1}^{d}( \widehat{\mathfrak{M}}_{\theta })
_{jk}\left\vert \psi _{j}\right\rangle \left\langle \psi _{k}\right\vert -2%
\frac{\sum_{k=1}^{m}p_{m}\left[ \partial _{\theta }p_{m}\right] }{%
	\sum_{k=1}^{m}p_{k}^{2}}.
\end{equation}%
Note that when $k>m$ and $j>m$ are satisfied simultaneously, $( \widehat{\mathfrak{M}}%
_{\theta }) _{kj}$ can take any value (for convenience, they can be
selected as $0$), otherwise $( \widehat{\mathfrak{M}}_{\theta }) _{kj}$ is
determined by Eq.~(\ref{e99}). Therefore, for a non-fully rank density matrix $\widehat{\rho}_{\theta}$, the form of $\widetilde{\mathfrak{M}}_{\theta } $ is also not unique, but this does not affect the value of DQFI.
\end{widetext} 
\section*{References}
\bibliography{Ref}

\end{document}